\documentclass[%
twocolumn,
 aps, 
prb,
floatfix,
]{revtex4-2}

\usepackage{dcolumn}
\usepackage{bm}
\usepackage[normalem]{ulem}
\usepackage{amsmath}
\usepackage{comment}
\usepackage{glossaries}
\usepackage{graphicx}
\usepackage{siunitx}
\usepackage{tabularx}
\usepackage{bm}
\usepackage{braket}
\usepackage[version=4]{mhchem}
\DeclareSIUnit\angstrom{\text {Å}}
\DeclareSIUnit\angstromcube{\text {Å$^3$}}

\setacronymstyle{long-short}
\newacronym{dft}{DFT}{density functional theory}
\newacronym{dho}{DHO}{damped harmonic oscillator}
\newacronym{fc}{FC}{force constant}
\newacronym{ahc}{AHC}{anomalous Hall conductivity}
\newacronym{anc}{ANC}{anomalous Nernst conductivity}
\newacronym{mca}{MCA}{magnetocrystalline anisotropy}
\newacronym{cfim}{CFiM}{compensated ferrimagnet}
\newacronym{fim}{FiM}{ferrimagnet}
\newacronym{htp}{HTP}{high-throughput}
\newacronym{oqmd}{OQMD}{Open Quantum Materials Database}
\newacronym{vasp}{VASP}{Vienna \textit{ab initio} Simulation Package}
\newacronym{sp}{SP}{spin polarization}
\newacronym{tc}{$T_{\mathrm{c}}$}{magnetic critical temperature}
\newacronym{bz}{BZ}{Brillouin zone}
\newacronym{pmg}{pymatgen}{Python Materials Genomics}
\newacronym{dos}{DOS}{density of states}
\newacronym{ase}{ASE}{Atomic Simulation Environment}
\newacronym{ifc}{IFC}{interatomic force constant}
\newacronym{asa}{ASA}{atomic sphere approximation}
\newacronym{fp}{FP}{full-potential}
\newacronym{icsd}{ICSD}{Inorganic Crystal Structure Database}

\begin{document}

\preprint{\it{preprint}}

\title{\textbf{High-throughput computational screening of Heusler compounds with phonon considerations for enhanced material discovery}
}%

\author{Enda Xiao}
\email{Xiao.Enda@nims.go.jp}
\affiliation{Research Center for Magnetic and Spintronic Materials, National
	Institute for Materials Science, 1-2-1 Sengen, Tsukuba, 305-0047, Ibaraki, Japan}
\author{Terumasa Tadano}
\email{Tadano.Terumasa@nims.go.jp}
\affiliation{Research Center for Magnetic and Spintronic Materials, National
	Institute for Materials Science, 1-2-1 Sengen, Tsukuba, 305-0047, Ibaraki, Japan}
\affiliation{Digital Transformation Initiative Center for Magnetic Materials (DXMag),
	National Institute for Materials Science, 1-2-1 Sengen, Tsukuba, 305-0047, Ibaraki, Japan}

\date{\today}

\begin{abstract}
	High-throughput (HTP) \textit{ab initio} calculations are performed on
	27,865 Heusler compositions, covering a broad range of regular, inverse, and
	half-Heusler compounds in both cubic and tetragonal phases. In addition to
	conventional stability metrics, such as formation energy, Hull distance, and
	magnetic critical temperature $T_{\mathrm{c}}$, phonon stability is assessed by
	systematically conducting \textit{ab initio} phonon calculations for over
	8,000 compounds. The performance of \textit{ab initio} stability criteria is
	systematically assessed against 189 experimentally synthesized compounds,
	and magnetic critical temperature calculations are validated using 59
	experimental data points. As a result, we identify 631 stable compounds as
	promising candidates for further functional material exploration. Notably,
	47 low-moment ferrimagnets are identified, with their spin polarization and
	anomalous Hall/Nernst conductivity calculated to provide insights into
	potential applications in spintronics and energy harvesting. Furthermore,
	our analyses reveal linear relationship between $T_{\mathrm{c}}$ and magnetization in 14
	systems and correlations between stability and atomic properties such as
	atomic radius and ionization energy. The regular/inverse structures
	preference in $X_2YZ$ compound and tetragonal distortion are also
	investigated for a broad Heusler family.
\end{abstract}

\maketitle


\section{Introduction}

Heusler alloys are renowned for their exceptional magnetic and functional
properties, including high-saturation magnetization, substantial \gls{mca},
elevated \gls{tc}, significant magnetocaloric effects, and notable
thermoelectric performance \cite{zhangHighthroughputDesignMagnetic2021}. The
unique combination of these properties, along with the diverse compositions
within the Heusler family, has prompted extensive \gls{htp} studies to explore
their potential for various applications
\cite{gillessenCombinatorialStudyFull2009,gillessenCombinatorialStudyInverse2010,
	carreteFindingUnprecedentedlyLowThermalConductivity2014,
	faleevHeuslerCompoundsPerpendicular2017,sanvitoAcceleratedDiscoveryNew2017,
	gaoHighthroughputScreeningSpingapless2019,
	maratheExplorationAll3dHeusler2023,nokyGiantAnomalousHall2020,marathe_exploration_2023,
	xingChemicalsubstitutiondrivenGiantAnomalous2024}. Previous \gls{htp}
investigations have primarily relied on stable compounds in various
databases, such as the Heusler database at the University of Alabama, the
\gls{oqmd}, and the AFLOW
database~\cite{nokyGiantAnomalousHall2020,kirklinOpenQuantumMaterials2015,estersAfloworgWebEcosystem2023}.
However, these databases are limited in the scope of the Heusler compound
family, restricting the exploration of functional material candidates.

In \gls{htp} studies, the initial candidate pool is typically narrowed by
assessing the thermodynamic stability of compounds. Thermodynamic stability is
commonly evaluated using formation energy and distance to the convex hull, which
quantify stability relative to its decomposition into constituent elements or
competing phases. However, the consideration of dynamical stability, ensuring
that a compound does not undergo structural phase transitions, is rarely
incorporated into \gls{htp} frameworks due to the computational cost and
complexity of phonon calculations in magnetic systems.

Since most of the intriguing properties of Heusler compounds are related to
magnetism, it is vital to ensure the thermal stability of their magnetic
configurations at application temperatures. This stability is typically assessed
using \gls{tc}. Previous studies have estimated \gls{tc} using linear
interpolation derived from experimental data
\cite{sanvitoAcceleratedDiscoveryNew2017,enamullahHighefficientDefectTolerant2018,
	wurmehlGeometricElectronicMagnetic2005,chenInitioPredictionHalfmetallic2006}.
Some efforts have also employed mean-field theory combined with exchange
interaction constants from first-principles calculations to estimate \gls{tc}
\cite{balluffHighthroughputScreeningAntiferromagnetic2017a,huHighthroughputDesignCobased2023}.
However, the reliability and accuracy of \gls{tc} calculations have not been
thoroughly investigated against experimental results in a comprehensive Heusler space.

The diversity of Heusler compounds arises not only from their composition but
also from the possible structures they can adopt for a given composition.
Heusler compounds with the $X_2YZ$ composition can typically crystallize in
either regular or inverse structures, assuming no chemical disorder. An
empirical rule of structural preference in $X_2YZ$ Heusler compounds, referred
to as Burch's rule in some studies, has been employed in various works
\cite{burchHyperfineStudiesSite1974,gorausMagneticPropertiesV2MnGa2023,
	gorausMagneticPropertiesTi2MnAl2020,kreinerNewMn2basedHeusler2014a,
	huHighthroughputDesignCobased2023}. While this rule has slightly different
descriptions in different works, the key point is that the inverse structure is
preferred when the $Y$ element is located to the right of the $X$ element in the
periodic table, and vice versa. This empirical rule has been validated by
\gls{htp} \textit{ab initio} studies containing a few hundred
compounds~\cite{kreinerNewMn2basedHeusler2014a,huHighthroughputDesignCobased2023},
but its validity for a broader compositional space is not clarified. Besides,
Heusler compounds typically crystallize in the cubic or tetragonal phase. The
tetragonal phase, also referred to as tetragonal distortion, is essential for a
material to exhibit \gls{mca}, which is particularly critical for developing
materials with strong perpendicular magnetic anisotropy (PMA)
\cite{winterlik_design_2012}.

Recently, Heusler alloys have also emerged as promising candidates for
\gls{cfim} systems. Compared to traditional ferromagnetic (FM) materials,
\gls{cfim} offers several advantages for spintronics applications due to its low
moment, such as faster switching speeds, higher storage densities, and greater
resistance to external magnetic fields.
\cite{finleySpintronicsCompensatedFerrimagnets2020}. Notably, \gls{cfim} systems
have inequivalent magnetic sublattices, enabling conventional electrical reading
and writing mechanisms on FM systems, such as anomalous Hall effect (AHE),
tunnel magnetoresistance (TMR), spin-transfer torque (STT), and spin-orbit
torque (SOT) \cite{finleySpintronicsCompensatedFerrimagnets2020}. However, only
a few Mn-contained \gls{cfim} Heusler compounds have been synthesized and
investigated \cite{ kurtCubicMn2GaThin2014, thiyagarajahGiantSpontaneousHall2015,
	bettoSitespecificMagnetismHalfmetallic2015a,
	nayakDesignCompensatedFerrimagnetic2015,
	sahooCompensatedFerrimagneticTetragonal2016,
	stinshoffCompletelyCompensatedFerrimagnetism2017,
	gavreaInvestigationsCompensatedFerrimagnetism2020,
	chatterjeeEmergenceCompensatedFerrimagnetic2021,
	seredinaCompensatedFerrimagnetismCompensation2022,
	harikrishnanSpinSemimetallicBehavior2023}. \gls{cfim} can be obtained from
low-moment \gls{fim} by adjusting the chemical composition and concentration.
Thus, a low magnetic moment \gls{fim} are potential candidates for \gls{cfim},
but only a few systems have been proposed by
first-principles calculation~\cite{wurmehlValenceElectronRules2006,
	beneaHalfmetallicCompensatedFerrimagnetism2019,
	zhangPredictionFullyCompensated2019a,
	shiPredictionMagneticWeyl2018,nokyStrongAnomalousNernst2018}. Thus, an expanded
list of candidates taking stability into account is desired.

In this work, we conducted a \gls{htp} \textit{ab initio} search for stable
Heusler compounds covering a broad range of regular, inverse, and half structures
in both cubic and tetragonal phases, significantly expanding the pool of
materials available for functional exploration. Stability was assessed using
conventional metrics such as formation energy, hull distance, and magnetic
transition temperature. Additionally, we incorporated phonon stability, a
critical factor omitted in previous high-throughput investigations.
To confirm the validity of our screening and modeling procedure, we
benchmarked the proposed stability criteria against a dataset of 189 experimentally
synthesized compounds and the employed \gls{tc} calculation methods against 59 experimental
data points. This comprehensive analysis identified 631 compounds that
satisfy all stability criteria, marking them as promising candidates for further
functional material exploration. Notably, we identified 47 low-moment \gls{fim}
systems that meet all stability criteria. For these systems, we calculated the
\gls{sp}, \gls{ahc}, and \gls{anc}, providing insights into their potential
applications in spintronics and energy harvesting devices.

Our comprehensive \gls{htp} result also revealed significant linear relationships
between \gls{tc} and magnetization in 14 systems.
We also observed correlations
between compound stability and fundamental atomic properties such as atomic
radius and ionization energy. Additionally, we confirmed that inverse Heusler
compounds generally exhibit a lower electronegativity for the $X$ element compared
to the $Y$ element, alongside a comparable covalent radius difference between
these elements. Finally, we confirmed that tetragonal distortion correlates with
a high density of states at the Fermi level in the cubic phase for $X_2YZ$
compounds. We also found that this correlation extends to the studied half-Heusler compounds.

\section{Methodology}

\begin{figure*}[bhpt]
	\centering
	\includegraphics[width=0.98\textwidth]{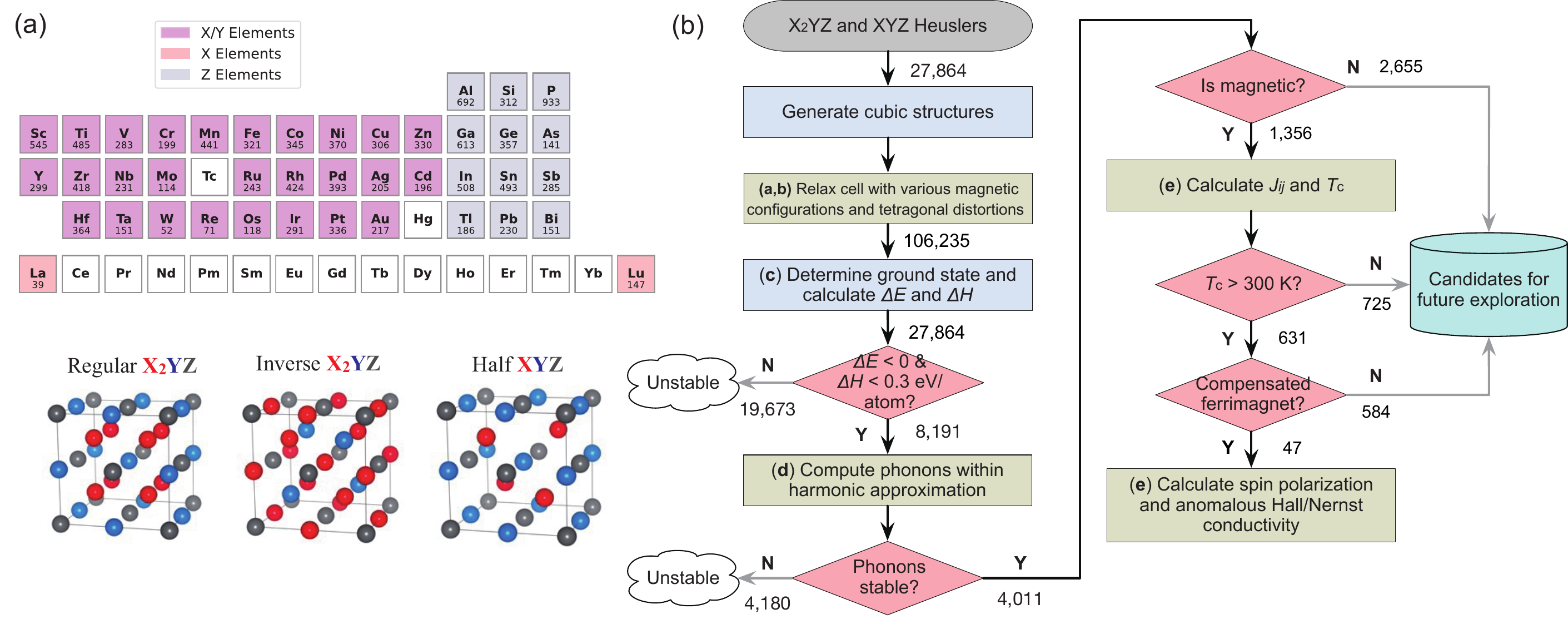}
	\caption{ (a) The compositions covered in the high-throughput search and
		distribution of stable compositions. Under each element, the number of
		compounds that contain this element and meet the stability criteria ($\Delta
			E < 0.0$~eV/atom and $\Delta H < 0.3$~eV/atom, and $\omega_{min}=0$) is shown.
		(b) Workflow of the high-throughput search for stable Heusler compounds.}
	\label{fig:schematic}
\end{figure*}

\subsection{Composition and structure}

This \gls{htp} study investigates an extensive set of conventional Heusler
compounds, as illustrated in Fig.~\ref{fig:schematic} (a). The regular and
inverse Heusler compounds share the $X_2YZ$ composition, while half-Heusler
compounds adopt the $XYZ$ composition. The regular and inverse structures are
identical when the $X$ and $Y$ elements are the same, resulting in an $X_3Z$
composition. We considered all combinations of elements where $X$ and $Y$ are
transition metals from the d-block (excluding Tc and Hg), and $Z$ is a main
group element from groups 13, 14, or 15 in the p-block, as depicted in
Fig.~\ref{fig:schematic} (a). For the $X$ element, we additionally considered La
and Lu, whose 4f orbitals are empty or fully occupied.  This comprehensive
screening resulted in a total of 27,864 compounds, including 9,072 regular,
9,072 inverse, 9,396 half-Heusler compounds, and 324 $X_3Z$ compounds.

For regular and inverse Heusler compounds, the cubic regular structure belongs
to the $\mathrm{Fm}\overline{3}\mathrm{m}$ space group, with the $X$, $Y$, and
$Z$ atoms occupying the $8\mathrm{c~}(\frac{1}{4},\frac{1}{4},\frac{1}{4})$,
$4\mathrm{b~}(\frac{1}{2}, \frac{1}{2}, \frac{1}{2})$, and $4\mathrm{a~}(0,0,0)$
Wyckoff positions, respectively. The cubic inverse structure belongs to the
$\mathrm{Fm}\overline{4}3\mathrm{m}$ space group, with the $X_1$, $X_2$, $Y$,
and $Z$ atoms at $4\mathrm{c~}(\frac{1}{4},\frac{1}{4},\frac{1}{4})$,
$4\mathrm{b~}(\frac{1}{2},\frac{1}{2},\frac{1}{2})$,
$4\mathrm{d~}(\frac{3}{4},\frac{3}{4},\frac{3}{4})$, and $4\mathrm{a~}(0,0,0)$
Wyckoff positions, respectively. Half-Heusler compounds with $XYZ$ composition
adopt the cubic $\mathrm{Fm}\overline{4}3\mathrm{m}$ space group, where the $X$,
$Y$, and $Z$ atoms occupy the
$4\mathrm{c~}(\frac{1}{4},\frac{1}{4},\frac{1}{4})$,
$4\mathrm{b~}(\frac{1}{2},\frac{1}{2},\frac{1}{2})$, and $4\mathrm{a~}(0,0,0)$
Wyckoff positions, respectively. The tetragonal variants of regular, inverse,
and half-Heusler structures belong to the $\mathrm{I4/mmm}$,
$\mathrm{I\overline{4}m2}$, and $\mathrm{I\overline{4}m2}$ space groups for
regular, inverse, and half structures, respectively.

\subsection{Workflow for high-throughput calculation}

The workflow of the high-throughput \textit{ab initio} calculation is
schematically shown in Fig.~\ref{fig:schematic}(b), which we elaborate below.

\paragraph{Cell parameter optimization} The initial structures of cubic
Heuslers $X_2YZ$ (or $XYZ$) were generated using a primitive cell containing one
formula unit, where the initial lattice constant, $a_{\mathrm{ini}}$, was set to
the value reported in the \gls{oqmd} for the same Heusler if available. When an
\gls{oqmd} calculated result was not available for $X_2YZ$ (or $XYZ$), the
corresponding $a_{\mathrm{ini}}$ value was estimated by simply taking the
average of the $a$ values of the \gls{oqmd}-calculated Heuslers $X_2YZ^{\prime}$
(or $XYZ^{\prime}$) that share the same $X$ and $Y$ elements. The cell
parameters were then optimized by using the
\gls{vasp}~\cite{kresseEfficiencyInitioTotal1996,kresseUltrasoftPseudopotentialsProjector1999,perdewGeneralizedGradientApproximation1996}.
For each composition, we considered various initial magnetic configurations
following Ref.~\cite{ma_computational_2018}. Specifically, for the regular
Heusler $X_2YZ$ and half Heusler $XYZ$, the $X$-site magnetic moment can either
be parallel or antiparallel to the $Y$-site magnetic moment. For each spin
configuration, we tested two magnitudes for the local magnetic moment, i.e.,
$|\bm{m}_{i}|=$ 1 and 4 $\mu_{\mathrm{B}}$, to explore potential high-spin and
low-spin states. Hence, four initial magnetic configurations were considered. In
the case of the inverse Heusler $X_2YZ$, the spin moments at the two $X$ sites
can be antiparallel, so we considered eight initial configurations in total.
After the structure optimizations were finished, the final energies and magnetic
moments were compared to identify inequivalent (meta) stable states.
\paragraph{Tetragonal distortion} For each inequivalent (meta) stable
magnetic state identified, the cubic primitive cell was converted into a
conventional cell containing two formula units, and the $c$-axis length was
changed from its original value by $\pm$2\%, $\pm$10\%, $+30$\%, and $+50$\%.
The cell parameters were then optimized further from these initial structures 
using \gls{vasp} to explore potential lower-energy tetragonal phases. After the
optimization, the structure with the lowest energy was identified
as the ground state for the composition $X_2YZ$ (or $XYZ$).

\paragraph{Thermodynamic stability} For a compound to be stable against
decomposition into its constituent elements, the formation energy ($\Delta E$)
must be negative. In addition, a compound is considered thermodynamically stable
if its energy is lower than that of all possible competing phases. This relative
stability can be evaluated by the distance to the convex hull ($\Delta H$), and
only compounds on the convex hull ($\Delta H=0$) are thermodynamically stable in
the strict sense. However, it is possible that metastable phases ($H>0$) at zero
Kelvin become most stable at finite temperatures or under external strain.
Indeed, previous studies reported that many metastable compounds had been
synthesized experimentally as long as $H$ is reasonably small~\cite{Sun2016-nv}. Hence, we relax
the criterion and assume compounds to be thermodynamically stable if $\Delta H <
	0.3$~eV/atom, which was chosen based on the analysis in the previous work. This
choice is reasonable also for the Heuslers, as we will demonstrate below. To
summarize, compounds satisfying both $\Delta E < 0.0$~eV/atom and $\Delta H <
	0.3$~eV/atom were deemed thermodynamically stable and subject to dynamical
stability analysis.

\paragraph{Dynamical stability}
For assessing the dynamical stability of Heuslers, phonon calculations were conducted within
harmonic approximation using the ALAMODE package
\cite{tadanoSelfconsistentPhononCalculations2015,
	tadanoAnharmonicForceConstants2014}. The dynamical stability was assessed based
on the presence/absence of unstable phonon modes on the $\bm{q}$ points 
commensurate with the supercell size. In this work, a $2\sqrt{2}$ $\times$
$2\sqrt{2}$ $\times$ 2 conventional cell containing 32 formula units was
employed; hence, there are 32 commensurate $\bm{q}$ points in the first
\gls{bz}. If any of the eigenvalues of the dynamical matrix, i.e., the squared
phonon frequencies $\{\omega_{\bm{q}\nu}^{2}\}$, were negative at the
commensurate $\bm{q}$ points, the compound was assessed to be dynamically
unstable. If all $\omega_{\bm{q}\nu}^{2}$ values were non-negative, the compound
was assumed to be dynamically stable and subject to the subsequent property
calculations.

\paragraph{Property calculations} The dynamically stable compounds were
further split into magnetic and nonmagnetic systems. If the absolute sum of
local magnetic moment per formula unit, $\sum_{i} |\bm{m}_{i}|$, is larger than
\SI{0.1}{\mu_B}, the system was considered magnetic. For the magnetic systems,
the \gls{tc} was evaluated using the spin-polarized relativistic
Korringa-Kohn-Rostoker (SPRKKR) code~\cite{ebertCalculatingCondensedMatter2011},
and compounds having \gls{tc} higher than 300 K were identified. Among them, we
identified compounds that simultaneously satisfy $\sum_{i}|\bm{m}_{i}|>0.5$
\si{\mu_B} and $|\sum_{i}\bm{m}_{i}|<0.5$ \si{\mu_B} as \gls{cfim} candidates,
for which the \gls{anc} and \gls{ahc} were computed systematically by using
Wannier90~\cite{mostofiWannier90ToolObtaining2008,
	nagaosaAnomalousHallEffect2010}.

\subsection{Computational methods}

\Gls{dft} calculations were mainly conducted using
\gls{vasp}~\cite{kresseEfficiencyInitioTotal1996, kresseEfficientIterativeSchemes1996}. The projector augmented wave
method and the generalized gradient approximation (GGA) with the
Perdew-Burke-Ernzerhof (PBE) functional were
used~\cite{kresseUltrasoftPseudopotentialsProjector1999,perdewGeneralizedGradientApproximation1996}.
A plane-wave energy cutoff of \SI{520}{eV} was applied, and the \gls{bz} was
sampled using an automatically generated $\bm{k}$-point mesh by setting
KSPACING=\SI{0.2}{\angstrom}$^{-1}$ except for the body-centered tetragonal
(bct) lattice; for the bct lattice, the $\bm{k}$-mesh was generated using the
\gls{pmg} library with a reciprocal density of \SI{450}{\angstromcube}. The
Methfessel-Paxton
smearing~\cite{methfesselHighprecisionSamplingBrillouinzone1989} with widths of
\SI{0.05}{eV} and \SI{0.1}{eV} was employed for structural optimization and
phonon calculations, respectively. For electron \gls{dos} calculations, the
primitive cell was used and $\bm{k}$-mesh was generated using \gls{pmg} with a
denser density of \SI{1000}{\angstromcube}, and the tetrahedron method with the
Bl\"ochl corrections was used~\cite{blochlImprovedTetrahedronMethod1994}.

The formation energy ($\Delta E$) was calculated as the energy difference
between the Heusler compound and its elemental constituents. The most stable
elemental reference structures were obtained from the \gls{oqmd} and then
relaxed by \gls{vasp} further using the \gls{dft} parameters described above to
obtain the energies of the elemental constituents. The $\Delta H$ values were
calculated using the formation energies computed for the Heuslers and those of
the competing phases obtained from the \gls{oqmd} (ver. 1.6). The \gls{htp}
optimization scripts utilized the \gls{pmg} package to generate input files for
different initial structures using recommended pseudopotentials
\cite{jainHighthroughputInfrastructureDensity2011,
	ongPythonMaterialsGenomics2013}. Also, the \gls{ase} toolkit \cite{ase-paper}
and the spglib \cite{Togo2024-xo} were employed for the structure file
manipulation and symmetry analysis, respectively.

The phonon calculations were performed based on the supercell approach, as
implemented in ALAMODE. Each atom in the supercell was displaced from its
equilibrium position by 0.01--0.02 \AA, and the atomic forces in the displaced
configurations were computed using \gls{vasp}. The second-order \glspl{ifc}
between the atoms in the supercell were fitted to the displacement-force
dataset, and the dynamical matrix was constructed from the \glspl{ifc} to obtain
the phonon frequencies. Phonon calculations for magnetic systems pose unique
challenges, as small atomic displacements can alter the magnetic configuration
in certain compounds. To circumvent this issue, we employed a two-step approach.
First, a static \gls{dft} calculation was performed for the supercell without
displacing atoms, and the resulting charge density was saved. Second, the
\gls{dft} calculations for the displaced configurations were performed using the
charge density obtained in the first step as initial values. Here, we also fixed
the total magnetic moment to the desired value, as implemented by the NUPDOWN
tag in \gls{vasp}. We confirmed that this approach resulted in consistent
magnetic moments for the displaced configurations in almost all cases. In-house
scripts were employed for high-throughput phonon computations.

The \gls{tc} values were determined within the mean-field approximation using
the exchange interaction constants
$J_{ij}$~\cite{andersonTheoryMagneticExchange1963}:
\begin{equation}
	T_C= \frac{2}{3 k_B} J_{\max },
\end{equation}
where $J_{\max }$ denotes the largest eigenvalue of the $J^{\mu \nu}$ matrix
defined as $J^{\mu \nu}=\sum_{j \in \nu} J_{0 j}$. Here, the index 0 is a fixed
site in sublattice $\mu$, and the summation runs over all sites in sublattice
$\nu$. $J_{ij}$ values were calculated using the Liechtenstein formula,
implemented in SPRKKR~\cite{liechtensteinLocalSpinDensity1987}. The
exchange-correlation effects were treated within the PBE functional. The
optimized structures by the VASP calculations were employed, and the local
magnetic moments $\{m_i\}$ from the VASP calculations were used as the initial
values. We confirmed overall consistency between the $\{m_i\}$ values obtained
using SPRKKR and VASP, as shown in Fig. S2 in Supplemental Materials. The $\bm{k}$-point mesh for the
self-consistent and $J_{ij}$ calculations was \(28 \times 28 \times 28\), and
the basis set NL was set to 4. The cutoff radius flag CLURAD was set to 5 in the
$J_{ij}$ calculation, with which we confirmed convergence in \gls{tc}.
Results employing the \gls{asa} or \gls{fp} method were
computed~\cite{huhneFullpotentialSpinpolarizedRelativistic1998}. The \gls{htp}
\gls{tc} calculation scripts utilized the ASE2SPRKKR package to generate input
files from the optimized
structures~\cite{ASE2SPRKKRSoftwarePackage,ISI:000175131400009}.

The anomalous Hall conductivity ($\sigma_{xy}$) was calculated using the Kubo
formula in terms of the Berry curvature, as implemented in the Wannier90:
\begin{equation}
	\sigma_{x y}=-\frac{e^2}{\hbar} \int_{\mathrm{BZ}} \frac{d \boldsymbol{k}}{(2 \pi)^3} \sum_n f\left(\epsilon_{n k}\right) \Omega_{n, x y}(\boldsymbol{k}),
\end{equation}
where $\hbar, e$, $\epsilon_{n k}$, $f$, and $\Omega_{n, x y}$ represent the
reduced Planck constant, positive elementary charge, eigenenergy, Fermi
distribution, and Berry curvature, respectively. The Berry curvature is defined
as
\begin{equation}
	\Omega_{n, x y}(\boldsymbol{k})=-2 \hbar^2 \operatorname{Im} \sum_{m(\neq n)} \frac{\langle n \boldsymbol{k}| \hat{v}_x|m \boldsymbol{k}\rangle\langle m \boldsymbol{k}| \hat{v}_y|n \boldsymbol{k}\rangle}{\left(\epsilon_{n \boldsymbol{k}}-\epsilon_{m \boldsymbol{k}}\right)^2},
\end{equation}
with $\hat{v}_x\left(\hat{v}_y\right)$ being the $k_x\left(k_y\right)$ component
of the velocity operator, and $|n k\rangle$ representing the eigenstate. From
the \gls{dft} band structure, Wannier functions were generated using the
selected columns of the density matrix (SCDM) method
\cite{damleDisentanglementEntanglementUnified2018,
	damleCompressedRepresentationKohn2015}. The anomalous Nernst conductivity
($\alpha_{xy}$)  at finite temperature $T$ was calculated from the
energy-dependent anomalous Hall conductivity
$\sigma_{xy}(\epsilon)$\cite{xiaoBerryPhaseEffectAnomalous2006,xiaoBerryPhaseEffects2010b}:
\begin{equation}
	\alpha_{x y} (T) =\frac{1}{e T} \int d \epsilon\;(\epsilon-\mu) \frac{\partial f (\epsilon, T)}{\partial \epsilon} \sigma_{x y}(\epsilon, T=0)
\end{equation}
where $\mu$ is chemical potential and set to Fermi energy in the calculation. To
realize integrations over the \gls{bz} in $\sigma_{xy}$ calculation, a
$\bm{k}$-mesh of $150 \times 150 \times 150$ points was used. The $\alpha_{xy}$
values were evaluated using $\sigma_{xy}(\epsilon)$ within a 0.8 eV energy
window around the Fermi energy.

\section{Results}

After relaxation, a total of 106,235 structures were identified, comprising
27,864 ground states and 78,371 metastable states. The stability of the ground
states was further evaluated. Among these ground states, 8,191 (29.4\%)
satisfied the thermodynamic stability criteria of $\Delta E < 0.0$~eV/atom and
$\Delta H < 0.3$~eV/atom. For the 8,191 screened compounds, we attempted
\textit{ab initio} phonon calculations and successfully obtained phonon
frequencies for 8,180 compounds. For the remaining 11 cases, the phonon
calculation results were not successful nor reliable either due to a convergence
issue in \gls{dft} calculations or a large fitting error in the force constant
estimation. Hence, these compounds were excluded from the subsequent
calculations. Out of the 8,180 phonon computed compounds, 4,011 (14.4\%) were
identified dynamically stable, which include 1,898 regular, 1,192 inverse, 81
$X_3Z$, and 840 half-Heusler compounds. These findings underscore the importance
of incorporating comprehensive stability metrics for efficient material
discovery, highlighting that phonon stability can significantly narrow the candidates' list. In
Fig.~\ref{fig:schematic}(a), the number of compounds that contain the
corresponding element and meet the stability criteria is shown under each
element.

\begin{table*}[h]
	\caption{List of stable ferrimagnets with a total magnetization smaller than
	0.5 $\mu_B$ and an absolute sum of local magnetic moment larger than 0.5
	$\mu_B$. The table provides the formation energy $\Delta E$ (eV/f.u.),
	distance to the convex hull $\Delta H$ (eV/f.u.), total magnetic moment
	$m_{\mathrm{tot}}$ ($\mu_B$), local magnetic moments $m_{\mathrm{loc}}$
	($\mu_B$), and magnetic critical temperature \gls{tc} (K) computed using
	the full-potential approach. Additionally, it includes the spin polarization
	SP, and electron density of states DOS (states/eV) at the Fermi level,
	anomalous Hall conductivity $\sigma_{xy}$ (S/cm), anomalous Nernst
	conductivity $\alpha_{xy}$ (A/Km), and the maximum anomalous Hall/Nernst
	conductivity $\sigma_{xy}^{\mathrm{max}}$/$\alpha_{xy}^{\mathrm{max}}$
	within an energy window of 250 meV around the Fermi level. In type column,
	r/i/h represents regular/inverse/half, and c/t represents cubic/tetragonal.}
	\label{tab:tab_cf}
	\resizebox{\textwidth}{!}{
		\begin{tabular}{llrrrllrrrrrr}
			\hline
			composition & type & $\Delta E$ & $\Delta H$ & $m_{\mathrm{tot}}$ & $m_{\mathrm{loc}}$           & \gls{tc} & SP     & DOS    & $\sigma_{xy}$ & $\sigma_{xy}^{\mathrm{max}}$ & $\alpha_{xy}$ & $\alpha_{xy}^{\mathrm{max}}$ \\
			\hline
			Cr$_2$AuAl  & it   & $-0.04$    & $0.16$     & $0.06$             & $[-2.90, 2.99, 0.01, -0.01]$ & $1932$   & $0.52$ & $3.28$ & $-658.02$     & $-718.50$                    & $0.42$        & $-1.78$                      \\
			Cr$_2$IrAl  & ic   & $-0.28$    & $0.19$     & $0.01$             & $[-2.05, 2.11, 0.00, 0.00]$  & $1310$   & $0.96$ & $6.10$ & $-325.17$     & $-603.50$                    & $-0.11$       & $2.37$                       \\
			Cr$_2$IrGe  & it   & $-0.10$    & $0.11$     & $0.29$             & $[2.53, -2.38, 0.12, 0.00]$  & $1292$   & $0.16$ & $6.08$ & $14.02$       & $663.00$                     & $1.51$        & $2.06$                       \\
			Cr$_2$MnAs  & rt   & $-0.03$    & $0.10$     & $0.03$             & $[-1.42, -1.42, 2.83, 0.08]$ & $760$    & $0.85$ & $4.81$ & $342.97$      & $689.58$                     & $0.83$        & $-2.30$                      \\
			Cr$_2$NiGe  & it   & $-0.05$    & $0.14$     & $0.01$             & $[2.29, -2.30, 0.00, 0.00]$  & $1557$   & $0.81$ & $5.14$ & $52.53$       & $199.49$                     & $-0.12$       & $1.00$                       \\
			Cr$_2$PdAl  & it   & $-0.16$    & $0.29$     & $0.10$             & $[2.88, -2.83, -0.01, 0.01]$ & $1802$   & $0.26$ & $4.89$ & $129.35$      & $385.71$                     & $0.13$        & $-1.19$                      \\
			Cr$_2$PdAs  & it   & $-0.03$    & $0.17$     & $0.21$             & $[3.04, -2.92, 0.05, 0.02]$  & $1518$   & $0.02$ & $3.91$ & $-86.87$      & $-390.94$                    & $0.42$        & $-1.35$                      \\
			Cr$_2$PdGa  & it   & $-0.11$    & $0.22$     & $0.10$             & $[2.93, -2.87, 0.00, 0.00]$  & $1813$   & $0.12$ & $5.02$ & $65.30$       & $66.14$                      & $0.21$        & $-0.51$                      \\
			Cr$_2$PdGe  & it   & $-0.08$    & $0.19$     & $0.04$             & $[-2.80, 2.85, 0.00, 0.00]$  & $1811$   & $0.68$ & $3.34$ & $-55.15$      & $-66.43$                     & $0.09$        & $-0.58$                      \\
			Cr$_2$PdSn  & it   & $-0.02$    & $0.27$     & $0.08$             & $[-3.07, 3.15, 0.01, 0.00]$  & $1759$   & $0.57$ & $2.94$ & $-18.76$      & $-69.87$                     & $0.19$        & $0.43$                       \\
			Cr$_2$PtAl  & it   & $-0.32$    & $0.20$     & $0.15$             & $[2.78, -2.71, 0.01, 0.00]$  & $1701$   & $0.31$ & $4.91$ & $235.79$      & $246.94$                     & $0.09$        & $-1.04$                      \\
			Cr$_2$PtGa  & it   & $-0.23$    & $0.10$     & $0.11$             & $[2.84, -2.78, 0.02, 0.00]$  & $1734$   & $0.19$ & $4.75$ & $100.26$      & $-266.89$                    & $0.09$        & $-1.26$                      \\
			Cr$_2$PtGe  & it   & $-0.16$    & $0.08$     & $0.05$             & $[-2.72, 2.80, -0.03, 0.00]$ & $1727$   & $0.63$ & $3.34$ & $-149.56$     & $-241.16$                    & $-0.57$       & $1.17$                       \\
			Cr$_2$PtIn  & it   & $-0.05$    & $0.21$     & $0.00$             & $[3.15, -3.16, -0.01, 0.00]$ & $1660$   & $0.23$ & $4.81$ & $-47.10$      & $176.52$                     & $0.11$        & $-1.01$                      \\
			Cr$_2$PtSi  & it   & $-0.27$    & $0.18$     & $0.01$             & $[2.46, -2.52, 0.04, 0.00]$  & $1633$   & $0.68$ & $4.02$ & $58.57$       & $321.61$                     & $0.34$        & $1.08$                       \\
			Cr$_2$PtSn  & it   & $-0.10$    & $0.21$     & $0.04$             & $[-3.01, 3.07, -0.02, 0.00]$ & $1754$   & $0.61$ & $2.83$ & $-181.73$     & $-186.59$                    & $0.12$        & $0.45$                       \\
			Cr$_2$RhGe  & it   & $-0.14$    & $0.14$     & $0.41$             & $[2.66, -2.42, 0.13, 0.00]$  & $1482$   & $0.12$ & $5.82$ & $-280.10$     & $-315.97$                    & $-1.20$       & $-1.81$                      \\
			Cr$_2$RhSb  & it   & $-0.04$    & $0.21$     & $0.03$             & $[-2.83, 2.92, -0.06, 0.01]$ & $1621$   & $0.67$ & $3.52$ & $9.60$        & $-136.89$                    & $0.14$        & $1.59$                       \\
			Cr$_2$RhSn  & it   & $-0.04$    & $0.23$     & $0.11$             & $[2.95, -2.92, 0.05, 0.00]$  & $1544$   & $0.32$ & $4.91$ & $-185.02$     & $-412.51$                    & $0.01$        & $2.03$                       \\
			CrMnAs      & hc   & $-0.04$    & $0.14$     & $0.00$             & $[-2.50, 2.60, 0.00]$        & $2244$   & $0.61$ & $2.82$ & $5.41$        & $202.19$                     & $0.32$        & $0.89$                       \\
			Mn$_2$AgAl  & ic   & $-0.02$    & $0.17$     & $0.33$             & $[-3.08, 3.38, 0.03, 0.01]$  & $1218$   & $0.56$ & $4.82$ & $220.09$      & $297.21$                     & $0.12$        & $-1.22$                      \\
			Mn$_2$AuIn  & it   & $-0.00$    & $0.12$     & $0.03$             & $[3.61, -3.60, 0.00, 0.01]$  & $782$    & $0.11$ & $5.45$ & $-25.94$      & $531.99$                     & $1.23$        & $1.48$                       \\
			Mn$_2$CuAl  & it   & $-0.18$    & $0.01$     & $0.20$             & $[-2.56, 2.77, 0.00, -0.01]$ & $1264$   & $0.60$ & $4.72$ & $11.41$       & $56.12$                      & $0.15$        & $0.16$                       \\
			Mn$_2$CuGa  & it   & $-0.11$    & $0.00$     & $0.35$             & $[-2.75, 3.06, 0.02, 0.00]$  & $1205$   & $0.57$ & $4.70$ & $11.09$       & $301.80$                     & $0.13$        & $1.03$                       \\
			Mn$_2$Ge    & ht   & $-0.04$    & $0.09$     & $0.00$             & $[-2.27, 2.36, -0.04]$       & $1751$   & $0.69$ & $3.28$ & $9.30$        & $119.51$                     & $-0.20$       & $-0.36$                      \\
			Mn$_2$IrAl  & it   & $-0.51$    & $0.00$     & $0.22$             & $[-2.81, 3.04, -0.01, 0.00]$ & $947$    & $0.38$ & $5.27$ & $-846.32$     & $-969.32$                    & $-0.91$       & $-1.44$                      \\
			Mn$_2$IrGa  & it   & $-0.39$    & $0.00$     & $0.20$             & $[-2.93, 3.13, -0.02, 0.01]$ & $984$    & $0.48$ & $5.12$ & $-836.28$     & $-881.77$                    & $-0.92$       & $-1.86$                      \\
			Mn$_2$IrIn  & it   & $-0.15$    & $0.01$     & $0.05$             & $[-3.32, 3.38, -0.02, 0.01]$ & $888$    & $0.64$ & $4.95$ & $-686.49$     & $-717.11$                    & $0.11$        & $0.94$                       \\
			Mn$_2$NiGe  & it   & $-0.22$    & $0.00$     & $0.37$             & $[-2.46, 2.75, 0.06, 0.00]$  & $917$    & $0.32$ & $5.02$ & $-14.88$      & $-97.78$                     & $-0.14$       & $-0.69$                      \\
			Mn$_2$NiSn  & it   & $-0.08$    & $0.07$     & $0.38$             & $[-3.00, 3.31, 0.04, 0.01]$  & $761$    & $0.40$ & $6.08$ & $167.06$      & $273.11$                     & $0.38$        & $0.81$                       \\
			Mn$_2$OsGe  & it   & $-0.13$    & $0.00$     & $0.03$             & $[2.74, -2.85, 0.16, -0.01]$ & $706$    & $0.89$ & $3.54$ & $194.50$      & $497.76$                     & $0.92$        & $-1.77$                      \\
			Mn$_2$PdIn  & it   & $-0.15$    & $0.11$     & $0.28$             & $[-3.46, 3.67, 0.04, 0.01]$  & $346$    & $0.21$ & $5.84$ & $33.27$       & $177.18$                     & $-0.01$       & $1.53$                       \\
			Mn$_2$PdSn  & it   & $-0.19$    & $0.10$     & $0.24$             & $[-3.37, 3.57, 0.02, 0.01]$  & $743$    & $0.13$ & $4.80$ & $22.92$       & $-141.62$                    & $0.18$        & $-0.52$                      \\
			Mn$_2$PtSn  & it   & $-0.23$    & $0.09$     & $0.04$             & $[3.48, -3.43, 0.00, 0.00]$  & $724$    & $0.11$ & $4.16$ & $153.43$      & $250.79$                     & $0.27$        & $1.28$                       \\
			Mn$_2$ReGe  & it   & $-0.07$    & $0.05$     & $0.22$             & $[2.48, -2.53, 0.27, -0.01]$ & $681$    & $0.31$ & $5.66$ & $-564.75$     & $-626.18$                    & $1.56$        & $-1.76$                      \\
			Mn$_2$RhAl  & it   & $-0.48$    & $0.07$     & $0.23$             & $[-2.90, 3.11, 0.01, 0.00]$  & $1066$   & $0.50$ & $5.64$ & $-510.42$     & $-658.36$                    & $-0.13$       & $1.54$                       \\
			Mn$_2$RhGa  & it   & $-0.40$    & $0.00$     & $0.16$             & $[-3.06, 3.21, 0.00, 0.01]$  & $1102$   & $0.56$ & $5.52$ & $-221.33$     & $-548.43$                    & $0.33$        & $1.12$                       \\
			Mn$_2$RhIn  & it   & $-0.19$    & $0.03$     & $0.01$             & $[-3.46, 3.46, 0.00, 0.01]$  & $996$    & $0.73$ & $4.81$ & $71.17$       & $518.14$                     & $-0.60$       & $-1.99$                      \\
			Mn$_2$RuGa  & it   & $-0.25$    & $0.00$     & $0.23$             & $[3.02, -3.01, 0.23, -0.02]$ & $1170$   & $0.00$ & $4.69$ & $-191.07$     & $377.05$                     & $-1.17$       & $-1.59$                      \\
			Mn$_2$RuIn  & it   & $-0.00$    & $0.05$     & $0.14$             & $[3.32, -3.35, 0.18, -0.02]$ & $1147$   & $0.13$ & $4.99$ & $-97.71$      & $-625.15$                    & $-1.08$       & $-1.78$                      \\
			Mn$_2$RuSb  & it   & $-0.07$    & $0.03$     & $0.20$             & $[-2.94, 3.09, 0.01, 0.01]$  & $568$    & $0.41$ & $6.35$ & $-357.75$     & $-692.35$                    & $-0.30$       & $1.14$                       \\
			Mn$_2$RuSn  & it   & $-0.11$    & $0.00$     & $0.02$             & $[3.13, -3.22, 0.13, -0.01]$ & $850$    & $0.80$ & $5.01$ & $278.54$      & $579.82$                     & $-1.72$       & $2.02$                       \\
			Mn$_2$Si    & hc   & $-0.19$    & $0.18$     & $0.01$             & $[1.29, -1.33, 0.02]$        & $1137$   & $0.67$ & $0.53$ & $-0.04$       & $-304.74$                    & $0.05$        & $-0.95$                      \\
			MnCrAs      & hc   & $-0.13$    & $0.05$     & $0.00$             & $[-1.45, 1.45, -0.02]$       & $1690$   & $0.87$ & $2.82$ & $125.80$      & $448.74$                     & $-0.16$       & $-1.05$                      \\
			Ti$_2$CrSn  & ic   & $-0.13$    & $0.19$     & $0.00$             & $[1.33, 0.98, -2.64, 0.02]$  & $1691$   & $0.96$ & $3.73$ & $56.23$       & $85.79$                      & $-0.09$       & $0.33$                       \\
			V$_2$ScGa   & ic   & $-0.05$    & $0.25$     & $0.30$             & $[-1.43, 1.57, 0.15, 0.00]$  & $585$    & $0.47$ & $9.21$ & $-468.32$     & $-468.32$                    & $-1.08$       & $-1.98$                      \\
			V$_2$TiSn   & ic   & $-0.08$    & $0.14$     & $0.04$             & $[1.42, -0.67, -0.69, 0.00]$ & $700$    & $0.22$ & $1.74$ & $-25.56$      & $88.31$                      & $0.07$        & $-0.35$                      \\
			\hline
		\end{tabular}}
\end{table*}

For magnetic applications, it is essential that compounds exhibit a magnetic
configuration that is stable at operational temperatures, typically room
temperature. Among thermodynamically and dynamically stable compounds, 1,356
compounds have been identified as magnetic satisfying $\sum_{i\in\mathrm{cell}}
	|\bm{m}_{i}|>0.1$ \si{\mu_B}. Of these, 631 compounds have a \gls{tc} exceeding
300~K, including 240 regular, 291 inverse, 24 $X_3Z$, and 76 half-Heusler
structures. The \gls{fp} \gls{tc} corrected by a factor of 0.85 were used. The
validation of \gls{tc} calculation method is included in subsection
\ref{sec:tc}. The full list of stable magnetic compounds is included in the
Supplemental Material.

Among the stable compounds, 47 low-moment \gls{fim} systems are identified,
characterized by a total magnetization $|\bm{m}_{\mathrm{tot}}|=|\sum_{i\in
		\mathrm{cell}}\bm{m}_{i}|$ smaller than 0.5 $\mu_B$ and $\sum_{i\in
		\mathrm{cell}}|\bm{m}_{i}|$ larger than 0.5 $\mu_B$. These systems are listed in
Table \ref{tab:tab_cf}. In previous work, it was proposed that compounds
containing 24/18 valence electrons for $X_2YZ$/$XYZ$ composition and Mn element
are promising candidates, and several such compounds were found by first-principle calculations
\cite{wurmehlValenceElectronRules2006,beneaHalfmetallicCompensatedFerrimagnetism2019,
	zhangPredictionFullyCompensated2019a}. Our finding expands the list of
low-moment Mn-containing compounds. We also find that Cr-containing compounds
\ce{Cr2YZ} ($Y$=Pd, Pt, Rh) can also exhibit low moments. Two V-containing
compounds and a Ti-containing compound are also identified. It should be noted
that the Slater-Pauling rule is not valid for some compounds, while the valence
electron number is close to the magic number 24 or 18. As suggested by the
appearance of \ce{MnCrAs} compound, we expect that quaternary type Heusler compounds
containing Mn and Cr can exhibit small $|\bm{m}_{\mathrm{tot}}|$ values,
potentially providing more \gls{cfim} candidates.

\begin{figure*}[bt]
	\centering
	\includegraphics[width=0.95\textwidth]{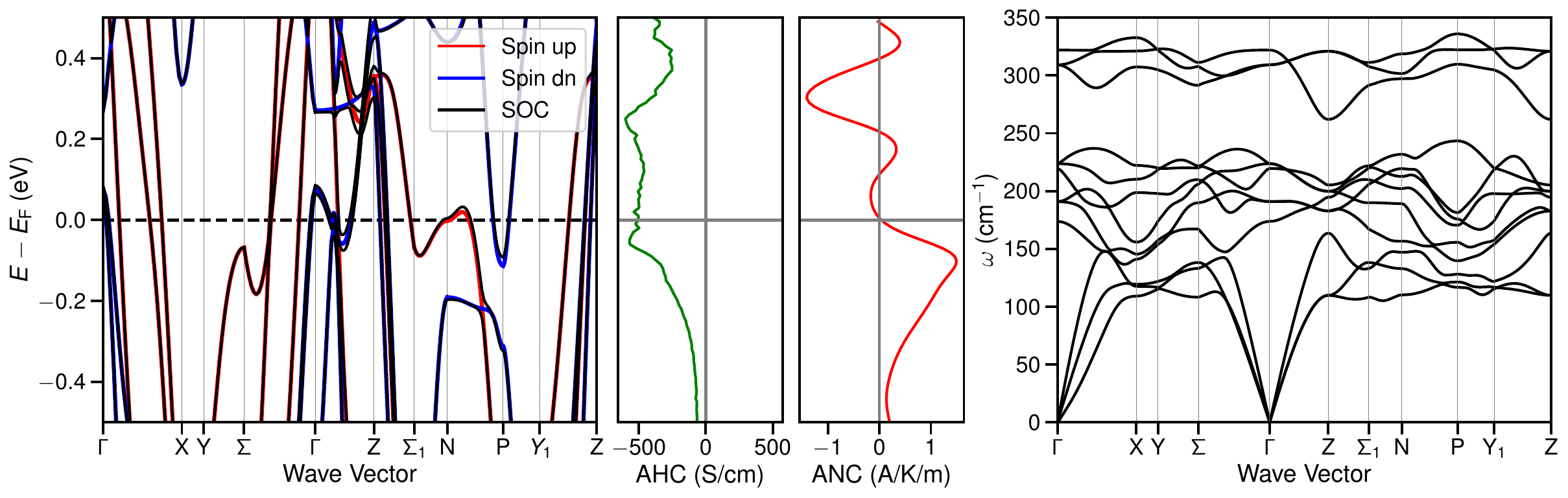}
	\caption{Calculated electronic band structures with and without spin-orbit
		coupling, \gls{ahc}, \gls{anc}, and the phonon band structure of inverse
		tetragonal Mn$_2$RhAl.}
	\label{fig:AHNC}
\end{figure*}

To aid in the discovery of functional
materials, we calculated the \gls{sp}, \gls{ahc}, and \gls{anc} to evaluate
their potential performance, as listed in Table \ref{tab:tab_cf}. Notably, 17
compounds exhibit a \gls{sp} greater than 0.6, which corresponds to a TMR of
112\% according to the Julli\`ere's formula
\cite{julliereTunnelingFerromagneticFilms1975}. Among these, \ce{Cr2MnAs}, \ce{MnCrAs},
\ce{Mn2OsGe}, \ce{Ti2CrSn}, and \ce{Cr2IrAl} stand out with the \gls{sp} values of
0.85, 0.87, 0.89, 0.96, and 0.96, respectively. In a recent work, low-moment
\gls{fim} compounds Ti$_2$Mn$Z$ ($Z$=Al, Ga) were reported to exhibit
promising AHC (253/268 \si{S.cm^{-1}}) and ANC at room temperature (1.31/0.94
\si{A.m^{-1}.K^{-1}}) by first-principle
calculation~\cite{shiPredictionMagneticWeyl2018, nokyStrongAnomalousNernst2018}.
Among the newly identified compounds, 12 compounds exhibit an AHC exceeding
\SI{250}{S.cm^{-1}} and 8 compounds exhibit an ANC exceeding
\SI{1}{A.m^{-1}K^{-1}}. Given that the Fermi level can shift with variations in
chemical composition and concentration in tuning for \gls{cfim}, we also present
the maximum AHC and ANC values within an energy window of 250 meV around the
Fermi level, as detailed in Table \ref{tab:tab_cf}.

As an example of the identified functional \gls{cfim} compounds, we show the
calculated electronic band structures with and without spin-orbit coupling, the
energy-dependent \gls{ahc}/\gls{anc}, and the phonon band structure of inverse
\ce{Mn2RhAl} in Fig.~\ref{fig:AHNC}. The \gls{ahc} shows a small variance around
the Fermi level, suggesting the potential that total magnetization can be tuned
by doping without lowering its \gls{ahc}. The structure is tetragonally
distorted with $a=5.43$ \si{\angstrom} and $c=7.30$ \si{\angstrom} whose local
magnetic moments are shown in Table \ref{tab:tab_cf}. This structure exhibits
$\Delta E=$ \SI{-0.48}{eV/atom} and $\Delta H=$ \SI{0.07}{eV/atom}, indicating
its potential synthesizability. The energy of the structure is
\SI{0.18}{eV/atom} lower than the most stable regular structure having the same
chemical formula \ce{Mn2RhAl}. Also, the $\Delta H$ value is lower than those of
the competing half Heuslers \ce{MnRhAl} and \ce{RhMnAl} by 0.60 and 0.43
\si{eV/atom}, respectively, suggesting the preferred stability in the inverse
structure. We also found another metastable inverse \ce{Mn2RhAl} that is less
stable than the identified \gls{cfim} candidate only by 0.006 \si{eV/atom}. This
competing structure is slightly distorted with $a=5.92$ \si{\angstrom} and
$c=5.94$ \si{\angstrom}, and its total magnetic moment is \SI{1.90}{\mu_B/cell},
which are consistent with the previously reported theoretical value for the
cubic inverse \ce{Mn2RhAl}~\cite{Ren2016-bb}. Since the energy difference
between the two inverse structures is small, experimental realization of the
\gls{cfim} candidate would require careful optimization of the synthesis
conditions.

\begin{figure}[tb]
	\centering
	\includegraphics[height=0.48\textwidth,clip]{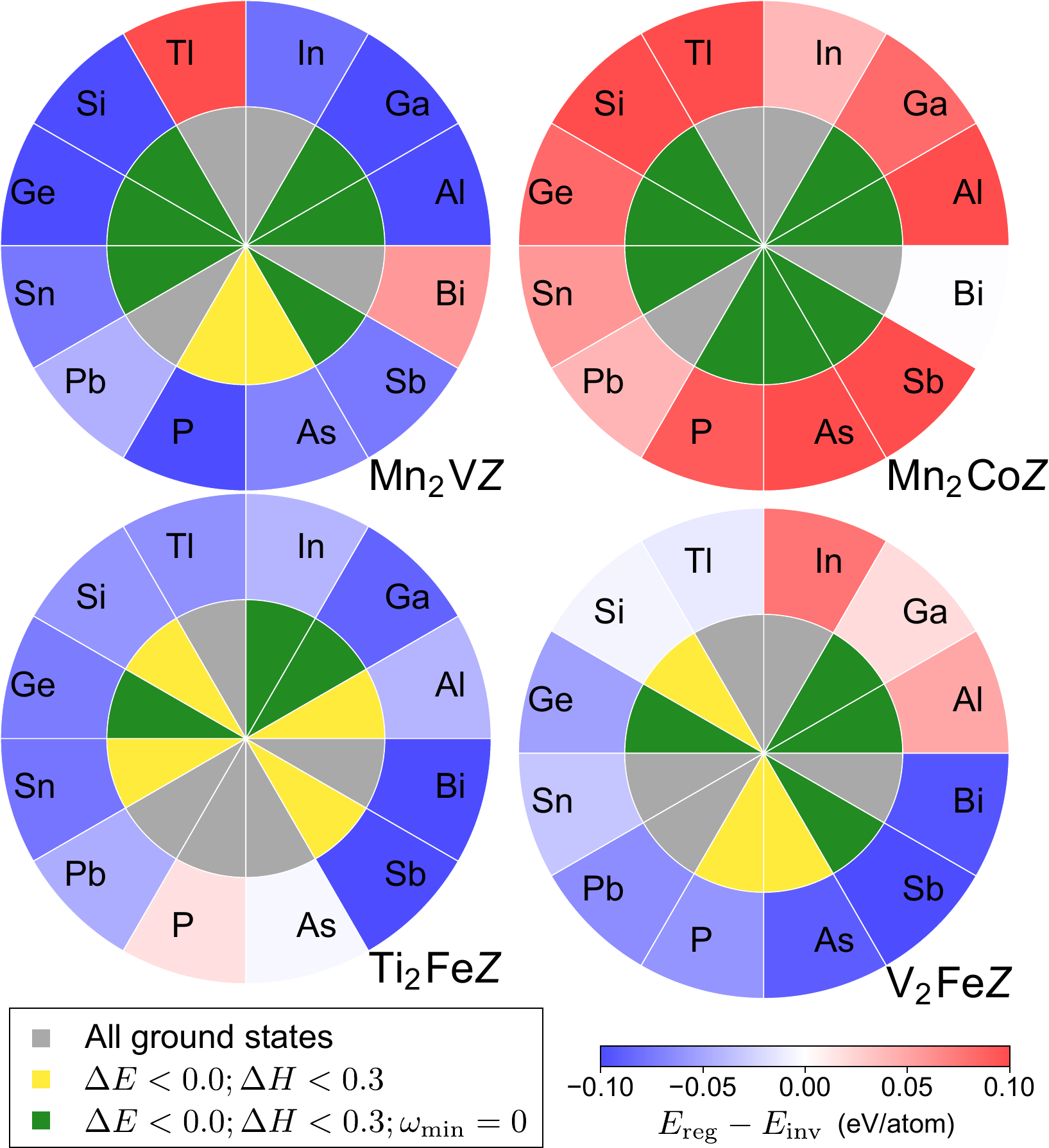}
	\caption{Structural preference and stability of Mn$_2$V$Z$, Mn$_2$Co$Z$,
		Ti$_2$Fe$Z$ and V$_2$Fe$Z$ compounds. The pie chart slices are arranged by
		the $Z$ element group. The inner circle indicates the stability categorized into three levels, while the outer-circle color map represents the energy difference $E_{\mathrm{reg}} - E_{\mathrm{inv}}$ in
		the range [\SI{-0.1}, \SI{0.1}]~eV/atom. Blue and red indicate preference
		for regular and inverse structures, respectively. Values outside this
		range are capped at the boundary values.}
	\label{fig:single_pie}
\end{figure}

Besides the discovery of promising candidates, the comprehensive \gls{htp}
dataset also provides insights into the stability and structural preference in
Heusler space. For example, \(X_2YZ\) has a preference for either the regular
structure or the inverse structure over the other. For visualizing the distribution
of regular and inverse structures and their relative stability, pie charts were
used, as shown in Fig.~\ref{fig:single_pie}, using Mn$_2$V$Z$,
Mn$_2$Co$Z$, Ti$_2$Fe$Z$, and V$_2$Fe$Z$ as examples. The pie chart slices are
arranged by the $Z$ element. The inner circle indicates stability, while the
outer-circle color map represents the energy difference $E_{\mathrm{reg}} -
	E_{\mathrm{inv}}$.  In the inner circle, the stability level is classified into
three categories based on $\Delta E$, $\Delta H$, and $\omega_{\mathrm{min}}$.
Here, $\omega_{\mathrm{min}}$ is the minimum value of the computed phonon
frequencies on the commensurate $\bm{q}$ points and is shown as a negative value
when it is imaginary. Hence, $\omega_{\mathrm{min}}=0$ holds when the compound
is dynamically stable. Given that the preference for regular or inverse
structures is significant when the energy difference exceeds 0.1 eV/atom, the
color map is capped at 0.1 eV/atom. In these examples, Mn$_2$VAl and Mn$_2$VGa
are both stable and prefer the regular structure, while Mn$_2$CoAl, Mn$_2$CoGa,
Mn$_2$CoGe, Mn$_2$CoSn, and Mn$_2$CoSb are stable and exhibit a preference for
the inverse structure, consistent with experimental
data~\cite{sanvitoAcceleratedDiscoveryNew2017}. The coexistence of regular and
inverse structures has also been experimentally observed in compounds such as
Ti$_2$FeAl, Ti$_2$FeGa, V$_2$FeAl, and
V$_2$FeGa~\cite{gorausMagneticPropertiesV2MnGa2023,
	gorausMagneticPropertiesTi2MnAl2020}. The pie charts in
Fig.~\ref{fig:single_pie} show the small energy differences ($E_{\mathrm{reg}} -
	E_{\mathrm{inv}}$) in these cases, which explain the coexistence of regular and
inverse structures.

\begin{figure*}[t]
	\centering
	\includegraphics[width=0.89\textwidth]{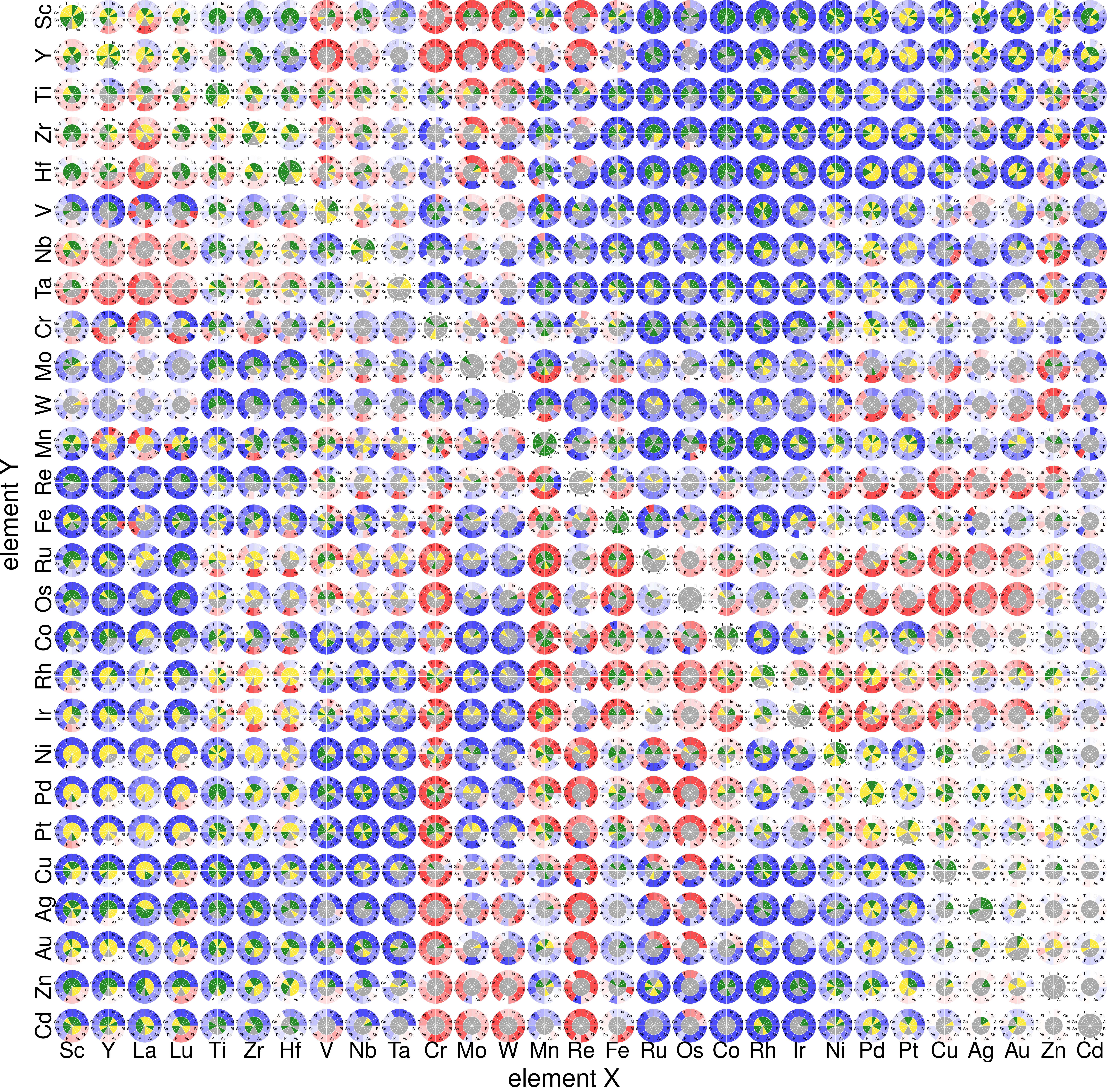}
	\caption{Comprehensive stability and structural preference map for $X_2YZ$
		Heusler compounds. The $x$ and $y$ axes represent the $X$ and $Y$
		elements, respectively. The order of elements is sorted by the group
		number for convenient comparison with empirical Burch's rule. Each pie
		chart corresponds to 12 compounds with the same $X$ and $Y$ elements but
		different $Z$ elements arranged in the same style as
		Fig.~\ref{fig:single_pie}. The inner circle indicates stability, while
		the outer circle represents $E_{\mathrm{reg}}- E_{\mathrm{inv}}$ in the
		range [\SI{-0.1}, \SI{0.1}]~eV/atom. Blue/red indicates a preference for
		regular/inverse structures. Values outside this range are capped.}
	\label{fig:pie}
\end{figure*}

Using this visualization style, we generated a comprehensive stability and
regular/inverse preference map for all $X_2YZ$ compounds, shown in
Fig.~\ref{fig:pie}. The $x$- and $y$-axes correspond to the $X$ and $Y$
elements, respectively, while each pie chart represents the stability and
structural preferences of 12 compounds with varying $Z$ elements. This
visualization style provides a clear and comprehensive overview of stability and
structural preferences over $X$, $Y$, and $Z$ compositions. Most stable inverse
compounds are distributed in the lower left corner of the map, aligning with the
empirical rule. However, a significant number of stable regular compounds are
also located in this region. A detailed discussion of the regular/inverse
preference is provided in subsection \ref{sec:reg_inv}.

We also generated maps for other competition
pairs. For compounds with the same $X$, $Y$, and $Z$ elements, the preference
between $X_2YZ$ and $XYZ$ compositions is determined by the Hull distance
difference between the half-Heusler and the lower energy regular or inverse
Heusler compound. For a given composition and atomic arrangement, the
competition between cubic and tetragonal phases is determined by the energy
difference between the two phases. Comprehensive maps illustrating the
competition between $X_2YZ$/$XYZ$ compositions and cubic/tetragonal phases are
provided in the Supplemental Material. These visualizations offer a clear and
comprehensive overview to aid further study in Heusler compounds space.

\section{Discussion}

\subsection{\Acrfull{tc}}
\label{sec:tc}

\begin{figure}[bt]
	\centering
	\includegraphics[width=0.46\textwidth]{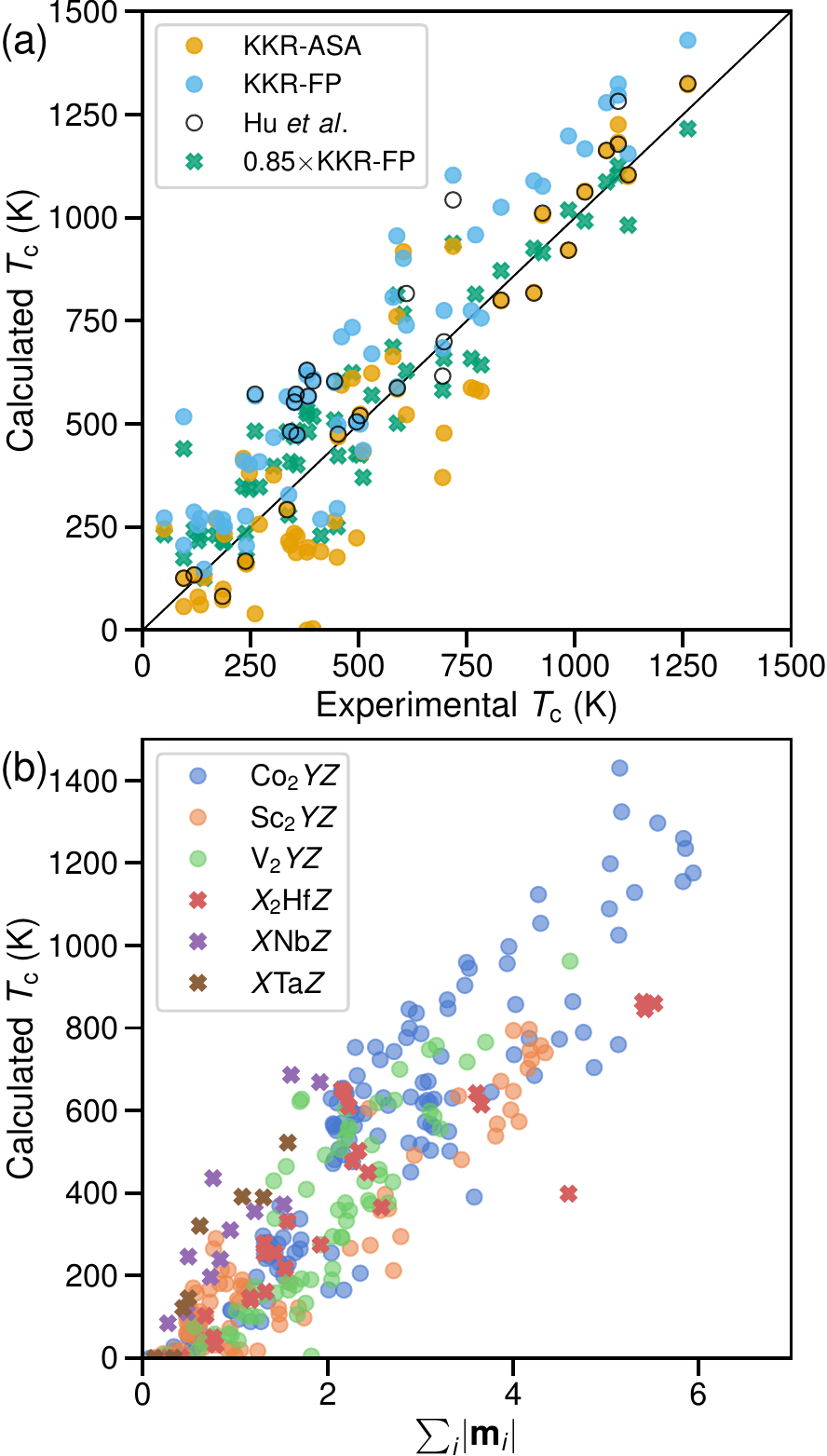}
	\caption{(a) Comparison of calculated and experimental \gls{tc}. Calculated
		\gls{tc} values using \gls{asa} and \gls{fp} methods are shown, along
		with FP results corrected by a factor of 0.85. Experimental data are from
		Refs.~\cite{sanvitoAcceleratedDiscoveryNew2017,huHighthroughputDesignCobased2023}.
		(b) The distribution of calculated \gls{tc} via \gls{fp} approach as a
		function of absolute sum of local magnetic moment $\sum_{i}
			|\bm{m}_{i}|$ in 6 Heusler compound systems.}
	\label{fig:Tc}
\end{figure}

The \gls{tc} values were computed within the mean-field approximation where the
exchange coupling constants ($J_{ij}$) were obtained using the magnetic force
theorem, as implemented in the SPRKKR code. The \gls{dft} calculations based on
the KKR method were performed either within the \gls{asa} or using the \gls{fp}
method. To assess the reliability of the computed \gls{tc} values, we first
compare the results with experimental data from previous high-throughput studies
by Sanvito \textit{et al.}~\cite{sanvitoAcceleratedDiscoveryNew2017} and Hu
\textit{et al.}~\cite{huHighthroughputDesignCobased2023}, as shown in
Fig.~\ref{fig:Tc} (a). Overall, the calculated \gls{tc} values show good
agreement with the experimental data. The \gls{tc} values obtained using the
\gls{asa} exhibit considerable scatter, sometimes exceeding and sometimes
falling below the experimental values. In contrast, the \gls{fp}-base \gls{tc}
values are consistently higher than the experimental data, which is reasonable
given the mean-field approximation's tendency to overestimate \gls{tc}
\cite{garaninSelfconsistentGaussianApproximation1996}. After applying a
correction factor 0.85 to the \gls{fp}-based \gls{tc} values, the results align
well with the experimental data, achieving an $R^2$ score of 0.87. This
adjustment compensates for the overestimation inherent in mean-field theory, and
the result is satisfactory for high-throughput screening purposes.

\begin{table}[htbp]
	\caption{Linear fit coefficients and $R^2$ scores for the correlation
		between \gls{tc} and the absolute sum of local magnetic moments,
		$\sum_{i} |\bm{m}_{i}|$. The linear relationship is expressed as
		\gls{tc}$= a \sum_{i} |\bm{m}_{i}| + b$. Results are presented for cases
		with data points number N larger than 10 and
		$R^2$ scores exceeding 0.7.}
	\label{tab:tab_Tc}
	\resizebox{0.48\textwidth}{!}{
		\begin{tabular}{llllllllll}
			\hline
			system     & N    & a     & b      & R$^2$  & system     & N     & a     & b      & R$^2$  \\
			\hline
			$X_2$Co$Z$ & $96$ & $217$ & $-145$ & $0.77$ & Lu$_2YZ$   & $33$  & $146$ & $-84$  & $0.85$ \\
			Sc$_2YZ$   & $79$ & $168$ & $-36$  & $0.87$ & Y$_2YZ$    & $31$  & $116$ & $-25$  & $0.85$ \\
			Ti$_2YZ$   & $64$ & $363$ & $-163$ & $0.71$ & $X_2$Hf$Z$ & $28$  & $155$ & $27$   & $0.78$ \\
			V$_2YZ$    & $64$ & $241$ & $-127$ & $0.72$ & $X_2$Nb$Z$ & $24$  & $164$ & $8$    & $0.79$ \\
			$X_2$Ti$Z$ & $60$ & $164$ & $-41$  & $0.73$ & Co$_2YZ$   & $132$ & $220$ & $-24$  & $0.83$ \\
			Co$YZ$     & $39$ & $252$ & $-69$  & $0.77$ & Fe$_2YZ$   & $125$ & $211$ & $-141$ & $0.73$ \\
			$X_2$Sc$Z$ & $37$ & $124$ & $38$   & $0.72$ & $X$Nb$Z$   & $12$  & $363$ & $-34$  & $0.80$ \\
			\hline
		\end{tabular}}
\end{table}

Additionally, a consistency check of the calculation method was performed by
comparing our calculated \gls{tc} values with those reported by Hu \textit{et
	al.} in Ref.~\cite{huHighthroughputDesignCobased2023}, which also utilized the
SPRKKR code and shown as a black empty circle in Fig. \ref{fig:Tc}. In
Ref.~\cite{huHighthroughputDesignCobased2023}, \gls{tc} values were primarily
calculated using the \gls{asa}, whereas the \gls{fp}-based computation was
conducted only when the magnetization from the \gls{asa} results was
inconsistent with the VASP magnetization. Consequently, some \gls{tc} values
align with our \gls{asa} results, while others align with our \gls{fp}-based
calculations. Additionally, there are four cases—Co$_2$MnAl, Co$_2$MnGa,
Mn$_2$CoGa, and Mn$_2$CoSn—where the results obtained by Hu \textit{et al.}
differ from both the ASA and \gls{fp} values in our study. These discrepancies
arise from differences in the identified ground states.

A previous study identified a linear relationship between the experimental
\gls{tc} and total magnetization in Co$_2$-based ferromagnetic Heusler compounds
across 10 different compositions \cite{wurmehlGeometricElectronicMagnetic2005}.
Our calculated \gls{tc} values, obtained via the \gls{fp} approach, exhibit the
same trend in certain systems. Fig. \ref{fig:Tc} (b) illustrates the
distribution of \gls{tc} values for six Heusler compound systems as a function
of the absolute sum of local magnetic moments, $\sum_{i} |\bm{m}_{i}|$. The
total magnetization is generalized to $\sum_{i} |\bm{m}_{i}|$ since there are
\gls{fim} and AFM compounds in our HTP results. Besides Co$_2YZ$ compounds,
this linear relationship is also observed in other Heusler systems, including
Sc$_2YZ$, V$_2YZ$, $X_2$Hf$Z$, $X$Nb$Z$, and $X$Ta$Z$, as shown in Fig.
\ref{fig:Tc}(b). Notably, while the linear trend holds across these systems, the
slope varies. For instance, the slope of $X$Nb$Z$ compounds is significantly
steeper than that of Sc$_2YZ$ compounds. Overall, we identified five systems
with an R$^2$ score exceeding 0.8 and nine additional systems with an R$^2$
score above 0.7. The corresponding linear fit coefficients and R$^2$ scores are
listed in Table \ref{tab:tab_Tc}. For systems with high R$^2$ values, this
linear relationship can be used to estimate the \gls{tc} of new compounds. The
distributions of \gls{tc} values versus $\sum_{i} |\bm{m}_{i}|$, categorized by
$X$ or $Y$ element species, are provided in Fig. S3 and S4 in the Supplemental
Material, along with the corresponding linear fit coefficients and R$^2$ scores.

The distribution of \acrlong{fim} (\acrshort{fim}) compounds among the 1,356
magnetic materials is noteworthy. Specifically, 12 (1\%) are antiferromagnetic,
and 636 (47\%) are \gls{fim}. All \gls{fim} compounds contain one or more of the
following elements: Mn, Cr, Fe, V, Co, or Ti. The counts of \gls{fim} compounds
containing Mn, Cr, Fe, V, Co, and Ti are 262, 123, 122, 120, 92, and 55,
respectively. In comparison, the counts of ferromagnetic compounds containing
these elements are 140, 41, 163, 47, 170, and 95, respectively. The ratio of
\gls{fim} ordering is high in compounds containing Mn, Cr, and V. This is in
agreement with a synthesis of such compounds in experiments
\cite{itohNeutronDiffractionStudy1983, chenDirectObservationFerrimagnetic2020,
jamerMagneticPropertiesLowmoment2016}. Detailed distributions of magnetic
ordering, categorized by element species and full/inverse/half Heusler
structures, are presented in Fig. S5 of the Supplemental Material.

\subsection{Refine stability criteria using experimental data}

\label{sec:criteria}

\begin{figure}[htbp]
	\centering
	\includegraphics[width=0.49\textwidth]{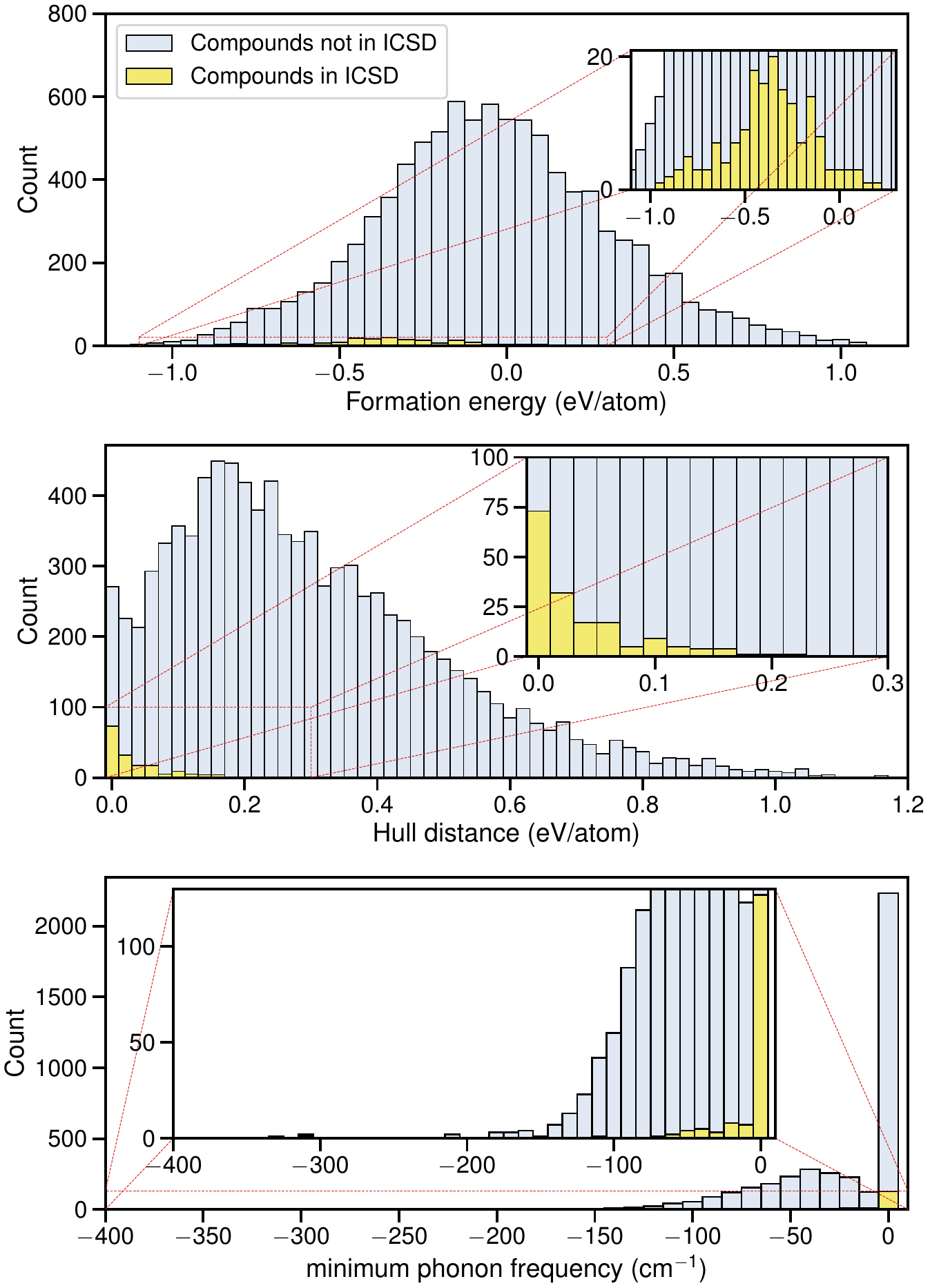}
	\caption{Distribution of $X_2YZ$ compounds ground states over formation
		energy, Hull distance, and minimum phonon frequency. Minimum phonon
		frequency distribution includes only compounds with $\Delta E <
			0.0$~eV/atom and $\Delta H < 0.3$~eV/atom. The number of compounds
		included in \gls{icsd} database are shown as yellow bins.}
	\label{fig:stability_distribution}
	\includegraphics[width=0.49\textwidth]{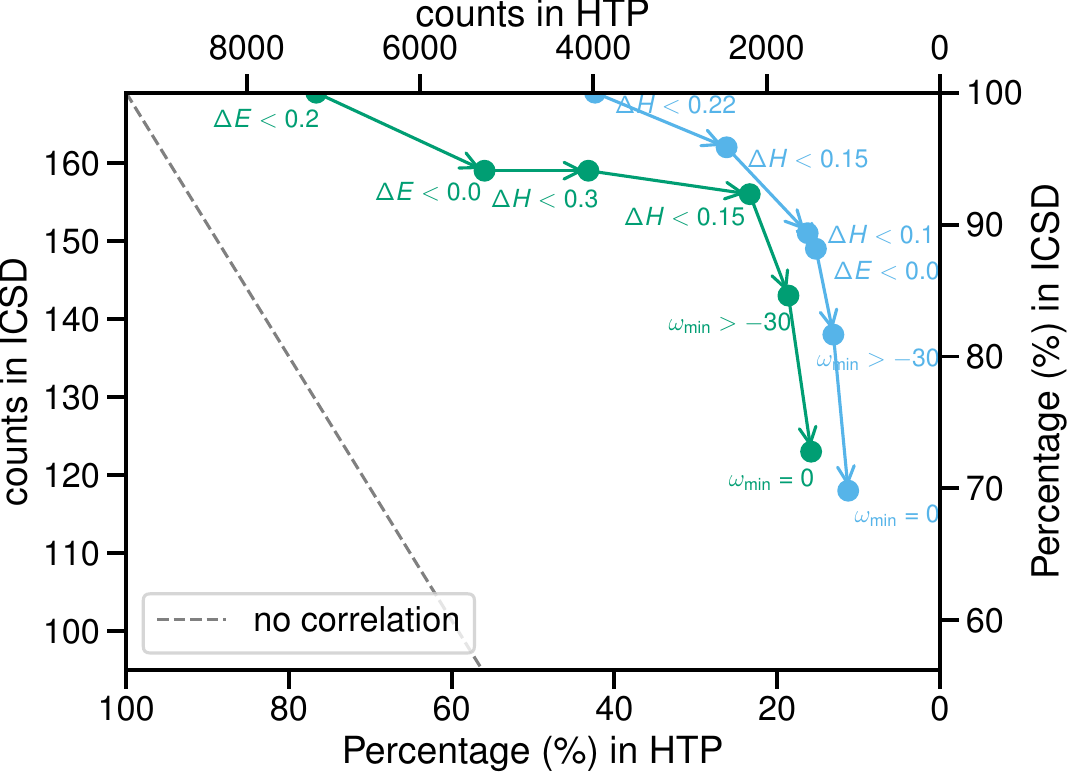}
	\caption{Performance of \textit{ab initio} stability criteria validated
		against \gls{icsd} experimental data. The $x$ axis shows the counts and
		percentages of \gls{htp} compounds satisfying sequentially applied
		stability criteria, while the $y$ axis indicates the counts and
		percentages of \gls{icsd} compounds included in the subset. If a
		criterion has no correlation with stability, it would fall on the
		diagonal line colored as gray. }
	\label{fig:stability_criteria}
\end{figure}

The stability of Heusler compounds was evaluated using three criteria: formation
energy ($\Delta E$), Hull distance ($\Delta H$), and minimum phonon frequency
($\omega_{\mathrm{min}}$). The distribution of $X_2YZ$ composition compounds
across these metrics is shown in Fig.~\ref{fig:stability_distribution}. For each
$X_2YZ$ composition, the structure (regular or inverse) with lower energy is
selected. Formation energy values are evenly distributed, peaking around
0~eV/atom. For minimum phonon frequency, a sharp peak at \SI{0}{cm^{-1}}
reflects that approximately half of the thermodynamically stable compounds are
also phonon-stable.

To assess the reliability of these stability criteria, we compare the stable
compound dataset from first-principles predictions to experimentally synthesized
compounds from the \acrfull{icsd}~\cite{belskyNewDevelopmentsInorganic2002}. The
\gls{icsd} contains 169 $X_2YZ$ and 21 $XYZ$ Heusler compounds. The
distributions of $\Delta E$, $\Delta H$, and $\omega_{\mathrm{min}}$ of the
\gls{icsd}-registered $X_2YZ$ compound are overlaid in
Fig.~\ref{fig:stability_distribution}. Additional figures for $XYZ$ compounds
are provided in the Supplemental Material.

Among the 169 $X_2YZ$ compounds from the \gls{icsd}, 159 (94\%) exhibit
formation energies below \SI{0}{eV/atom}, underscoring that negative formation
energy is a robust necessary condition for stability prediction with large
recall. A few \gls{icsd} compounds have positive calculated formation energies
of up to 0.2~eV/atom. Since formation energy is calculated at \SI{0}{K}, the
observed stability of these compounds in experiments might be due to the
stabilizing effect of entropic contributions at finite temperatures.
Additionally, the positive values could be attributed to uncertainties in
\gls{dft} calculations. In a compound search task prioritizing the
identification of as many candidates as possible, a relaxed threshold of $\Delta
	E$ is a reasonable choice.

The hull distance values for most \gls{icsd}-registered Heuslers are close to
zero. A few \gls{icsd} data extend up to $\Delta H = 0.22$ \si{eV/atom}, which
is lower than the $\Delta H < 0.3$ \si{eV/atom} threshold used in the \gls{htp}
workflow. Using a threshold of $\Delta H < 0.22$ \si{eV/atom} enhances precision
without compromising recall of \gls{icsd} data, while a stricter threshold of
$\Delta H < 0.10$ \si{eV/atom} can narrow down the candidate list by 62\%
compared to $\Delta H < 0.22$ \si{eV/atom} list, at the cost of a smaller recall
(89\%).

The distribution of $\omega_{\mathrm{min}}$ for compounds with $\Delta E < 0$
eV/atom and $\Delta H < 0.3$ eV/atom is presented in
Fig.~\ref{fig:stability_distribution}, including 159 compounds from the
\gls{icsd}. Imaginary phonon modes (represented as $\omega_{\mathrm{min}} < 0$)
indicate structural phase transitions and thus render the structures dynamically
unstable, as observed in compounds such as Pd$_2$TiSn, Pd$_2$ZrSn, and
Pd$_2$HfSn at low temperatures~\cite{suStructuralPhaseTransitions2022}. Among
the \gls{icsd} compounds, 124 (78\%) exhibit $\omega_{\mathrm{min}} = 0$,
affirming that the calculated phonon stability serves as a good criterion for
stability. The remaining 35 \gls{icsd} compounds display $\omega_{\mathrm{min}}$
values ranging from \numrange{-0.5}{-71.4} cm$^{-1}$ and two points at
\SI{-111.0}{cm^{-1}} and \SI{-308.9}{cm^{-1}}. The number of compounds with
negative $\omega_{\mathrm{min}}$ values decreases as the magnitude of
$\omega_{\mathrm{min}}$ becomes larger.

These predictions are based on phonon dispersion within the harmonic
approximation. It should be noted that this approach does not account for
anharmonic effects, which can stabilize systems with imaginary phonon modes at
finite temperatures~\cite{tadanoQuarticAnharmonicityRattlers2018,
	tadanoSelfconsistentPhononCalculations2015, suStructuralPhaseTransitions2022}.
The 35 \gls{icsd} compounds with negative $\omega_{\mathrm{min}}$ values may
become dynamically stable when anharmonic effects are included at finite
temperatures. However, due to the complexity and high computational cost
associated with treating anharmonic effects, a more practical strategy is to use
$\omega_{\mathrm{min}}$ computed at the harmonic level and choose a threshold
value at a balance of precision and recall based on the specific application.
For instance, a threshold of $\omega_{\mathrm{min}} = 0$ \si{cm^{-1}} is
recommended for tasks prioritizing reliable candidates, whereas a threshold of
$\omega_{\mathrm{min}} > -70$ \si{cm^{-1}} is suitable for identifying a broader
range of potential candidates.

As shown in Fig.~\ref{fig:stability_distribution}, applying stricter criteria
increases the precision of stability predictions, as the ratio of removed
compounds is larger in the \gls{htp} than the \gls{icsd} data, while recall is
reduced. Figure \ref{fig:stability_criteria} illustrates the number of the
\gls{htp} compounds that meet various stability criteria applied sequentially
and the number of the \gls{icsd}-registered Heuslers included within this
subset. The $x$-axis represents the numbers/percentages of \gls{htp} compounds
that satisfy various stability criteria applied sequentially. The number
decreases as the criteria become stricter. The $y$-axis represents the
numbers/percentages of \gls{icsd} compounds included in each set of \gls{htp}
compounds. The location of the points in the plot indicates the trade-off
between precision and recall. If a criterion has no correlation with stability,
it would fall on the gray diagonal line.  By switching the order in which
$\Delta E$ and $\Delta H$ are applied, it is clear that $\Delta H$ is a more
effective criterion for identifying thermodynamically stable compounds.

When using the criteria $\Delta E < 0.0$ eV/atom and $\Delta H < 0.3$ eV/atom,
4,058 (43\%) of \gls{htp} compounds are predicted to be stable, and 159 (94\%)
\gls{icsd} compounds are correctly identified as stable. When
$\omega_{\mathrm{min}} = 0$ cm$^{-1}$ is added as a criterion, 2,173 (23\%) of
\gls{htp} compounds are predicted to be stable, and 124 (78\%) \gls{icsd}
compounds are correctly identified as stable. This demonstrates the validity of
the stability criterion choice in our workflow.

From Fig. \ref{fig:stability_criteria}, we can identify optimal criteria for
different \gls{htp} screening task types quantitatively. In a screening task
prioritizing the identification of as many stable compounds as possible, a
relaxed threshold of $\Delta E$ is a reasonable choice. Using criteria $\Delta E
	< 0.2$ eV/atom and $\Delta H < 0.22$ eV/atom, 3,968 (42\%) of \gls{htp}
compounds are predicted to be stable, and all \gls{icsd} compounds are correctly
identified as stable. This set of criteria offers the best recall. Conversely,
with the criteria $\Delta E < 0.0$ eV/atom, $\Delta H < 0.10$ eV/atom, and
$\omega_{\mathrm{min}} = 0$ cm$^{-1}$, 1,057 (10\%) \gls{htp} compounds are
predicted to be stable, and 118 (70\%) \gls{icsd} compounds are correctly
identified as stable. This set of criteria is expected to provide good
precision, albeit with a trade-off in recall.

Similar trends are observed for half-Heusler compounds, where 1181 (13\%)
\gls{htp} compounds meet the criteria of $\Delta E < 0.0$~eV/atom and $\Delta H
	< 0.3$~eV/atom. Of 21 half-type \gls{icsd} compounds, 19 meet the same criteria.
The exception \ce{CuMnSb} has small $\Delta E = 0.05$~eV/atom and $\Delta H =
	0.07$~eV/atom. While the other exception \ce{CoCrAl} has large $\Delta H =
	0.44$~eV/atom, the reported experimental structure of \ce{CoCrAl} has Cr
occupying 50\% $(\frac{1}{2},\frac{1}{2},\frac{1}{2})$ and 50\%
$(\frac{3}{4},\frac{3}{4},\frac{3}{4})$ Wyckoff positions, which is not standard
half Heusler structure used in our computation
\cite{dattaMagneticPropertiesHeusler2022}. The stability observed in experiments
may be due to the disorder. When $\omega_{\mathrm{min}} = 0$ cm$^{-1}$ is added
as a criterion, 842 (9\%) \gls{htp} compounds are predicted to be stable, and 18
(85\%) \gls{icsd} compounds are correctly identified as stable.

\subsection{Stability and atomic feature}

\begin{figure}[bthp]
	\centering
	\includegraphics[width=0.48\textwidth]{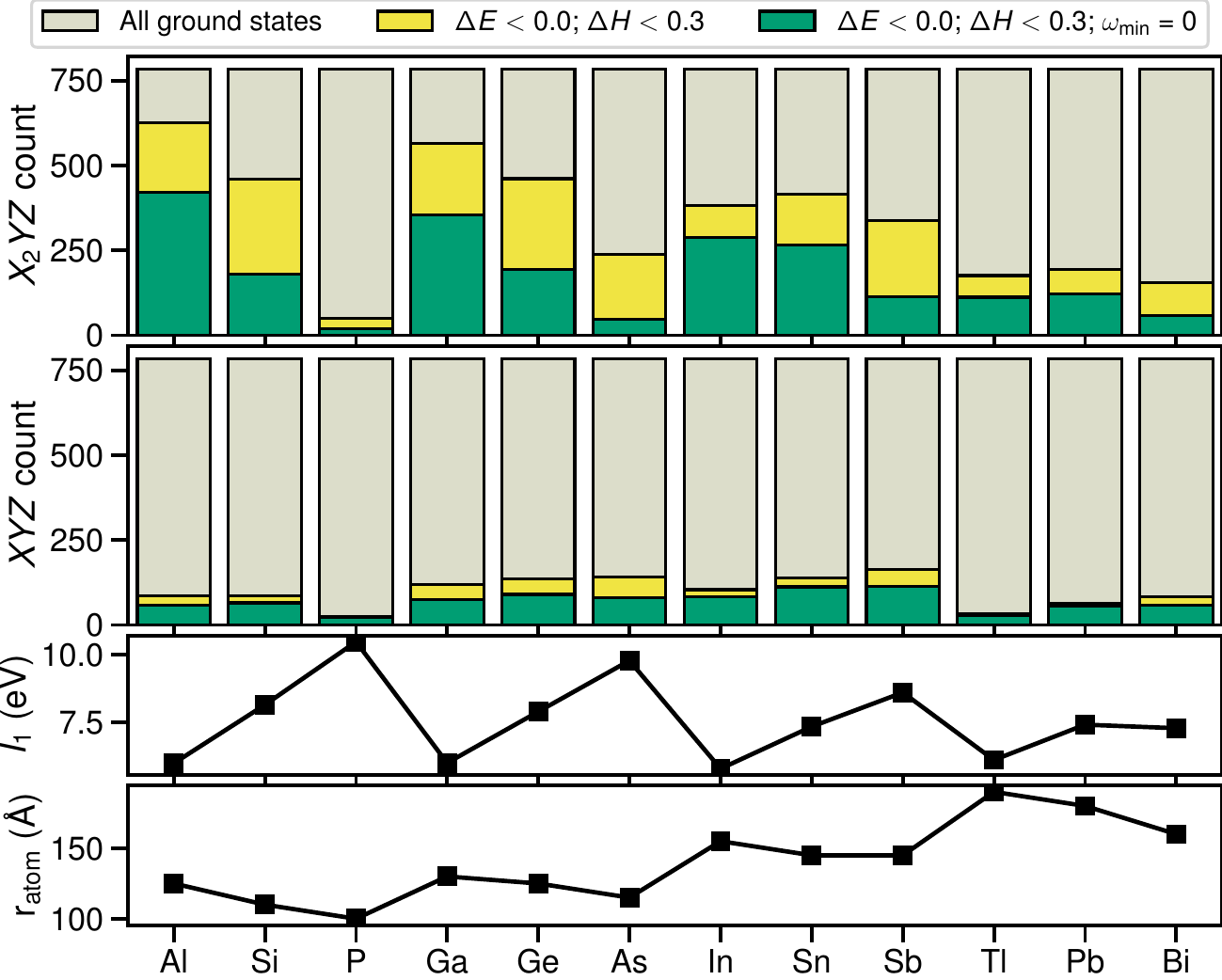}
	\caption{Distributions of $X_2YZ$ and half-type compounds over $Z$ element
		species, with different stability groups shown in various colors. First
		ionization energy ($I_1$) and atomic radius ($r_{\mathrm{atom}}$) of $Z$
		elements are sourced
		from~\cite{mendeleev2014,slaterAtomicRadiiCrystals1964,nist_asd}.}
	\label{fig:atom_X2YZ}
\end{figure}

From the \gls{htp} calculation results, a correlation between stability and
atomic properties is identified. For $X_2YZ$ compounds, stability is favored
when the $Z$ element has a small atomic radius and low ionization energy.
Figure~\ref{fig:atom_X2YZ} shows the distribution of $X_2YZ$ type compounds over
$Z$ element species, along with the $Z$ element's first ionization energy
($I_1$) and atomic radius ($r_{\mathrm{atom}}$)
\cite{mendeleev2014,slaterAtomicRadiiCrystals1964,nist_asd}. Each bin represents
the counts of compounds that contain the corresponding $Z$ element. The
compounds are divided into different stability groups indicated by various
colors. The local and global trends in stable compound distribution are
consistent with the trends in $I_1$ and $r_{\mathrm{atom}}$, respectively. For
comparison, the distribution of half-Heusler compounds is also shown, where no
such correlation is found. Among the phonon-stable $X_2YZ$ compounds, 51\%
contain either Al, Ga, or In as the Z element, and 26\% contain either Si, Ge, or
Sn. These elements should be prioritized for discovering stable functional
Heuslers. This observation is expected to be valid for more general Heusler
compounds, such as quaternary and off-stoichiometric types.

\begin{figure}[bthp]
	\includegraphics[width=0.49\textwidth]{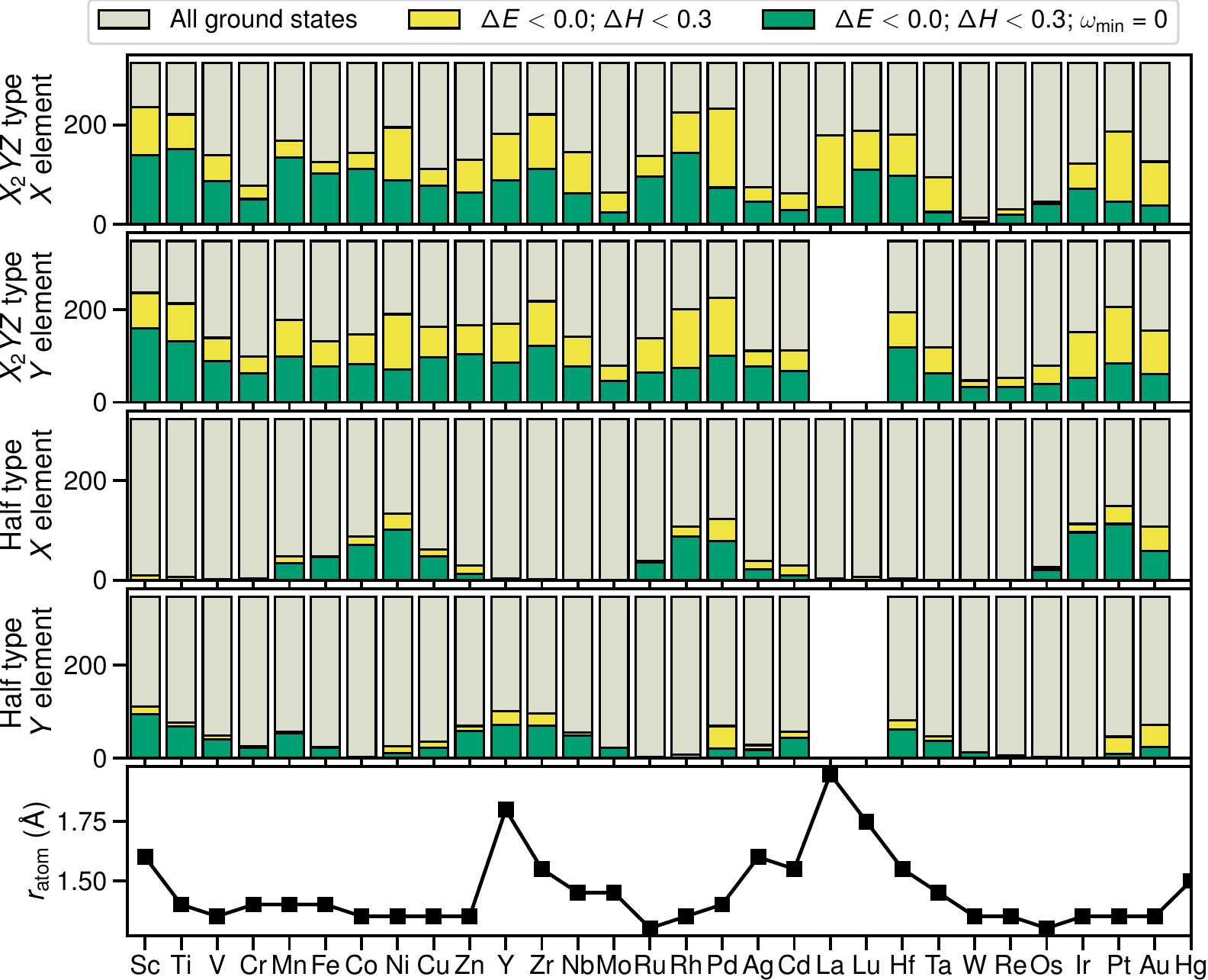}
	\caption{Distributions of $X_2YZ$ and half-type compounds over $X$ or Y
		element species, with different stability groups shown in various
		colors. Atomic radius ($r_{\mathrm{atom}}$) of $X$ and $Y$ elements are
		sourced from \cite{mendeleev2014,slaterAtomicRadiiCrystals1964}.}
	\label{fig:atom_half}
\end{figure}

In half-Heusler compounds, stability is higher when the $X$ element has a
smaller atomic radius compared to the $Y$ element. Figure \ref{fig:atom_half}
shows the distribution of half-Heusler compounds over $X$ and $Y$ element
species, along with the atomic radii of the $X$ and $Y$ elements. It is evident
that stable compounds are concentrated in the region where the $X$ element has a
smaller atomic radius, and the $Y$ element has a larger radius. For comparison,
the distribution of $X_2YZ$ compounds is also shown in Figure
\ref{fig:atom_half}, where no such correlation is observed. Notably, 94\% of
stable half-Heusler compounds satisfy the condition $r^X_{\mathrm{atom}} \leq
	r^Y_{\mathrm{atom}}$. This provides a robust necessary condition for selecting
stable half-Heusler candidates.

\subsection{Regular or inverse structure preference in $X_2YZ$ compound}
\label{sec:reg_inv}

\begin{figure*}[htbp]
	\centering
	\includegraphics[width=0.99\textwidth]{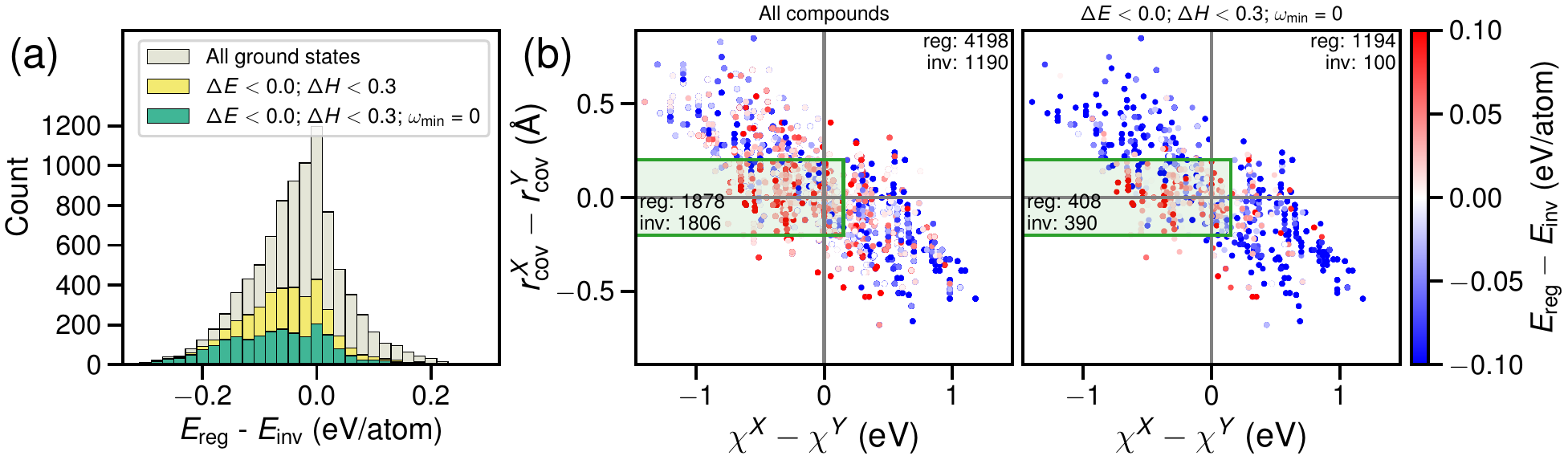}
	\caption{(a) Distribution of $X_2YZ$ compounds based on the energy
	difference between regular and inverse structures ($E_{\mathrm{reg}} -
		E_{\mathrm{inv}}$). Different stability groups are represented by various
	colors. (b) Regular/inverse preference of compounds meeting various
	stability criteria. The preference is shown by the energy difference
	($E_{\mathrm{reg}} - E_{\mathrm{inv}}$), which is color-coded. The
	electronegativity difference ($\chi^X - \chi^Y$) and covalent radius
	difference ($r^X_{\mathrm{cov}} - r^Y_{\mathrm{cov}}$) between $X$ and $Y$
	elements are shown on the $x$- and $y$-axes, respectively. The region where
	$\chi^X - \chi^Y < 0.15$ eV and $|r^X_{\mathrm{cov}} - r^Y_{\mathrm{cov}}| <
		0.20$ \si{\angstrom} is marked by green shading.}
	\label{fig:reg_inv}
\end{figure*}

Understanding the preference for regular or inverse structures in $X_2YZ$
compounds is crucial for tailoring their functional properties.
Figure~\ref{fig:reg_inv}(a) shows the distribution of $X_2YZ$ compounds based on
the energy difference between regular and inverse structures ($E_{\mathrm{reg}}
	- E_{\mathrm{inv}}$). Overall, the regular structure is generally preferred
across all $X_2YZ$ compounds analyzed. This trend remained consistent when
various stability criteria were applied.

The structural preference analysis for regular or inverse configurations in
$X_2YZ$ compounds is often guided by Burch's rule
\cite{burchHyperfineStudiesSite1974,gorausMagneticPropertiesV2MnGa2023,
	gorausMagneticPropertiesTi2MnAl2020,kreinerNewMn2basedHeusler2014a,
	huHighthroughputDesignCobased2023} According to this empirical rule, if the $Y$
element is positioned to the left of the $X$ element in the periodic table, the
compound is more likely to adopt a regular structure. Conversely, if the $Y$
element is to the right of the $X$ element, an inverse structure is favored.
This trend can be quantitatively represented using electronegativity differences
($\chi^X - \chi^Y$), as electronegativity generally increases from left to right
across the periodic table~\cite{huHighthroughputDesignCobased2023}. A negative
value of $\chi^X - \chi^Y$ indicates that $Y$ is to the right of $X$, favoring
an inverse structure.

This relationship is examined using our \gls{htp} calculation results and
illustrated in Fig.~\ref{fig:reg_inv}(b). The preference is shown by color-coded
energy difference ($E_{\mathrm{reg}} - E_{\mathrm{inv}}$), along with the
electronegativity difference ($\chi^X - \chi^Y$) and covalent radius difference
($r^X_{\mathrm{cov}} - r^Y_{\mathrm{cov}}$) between $X$ and $Y$ elements are
shown on the $x$- and $y$-axes, respectively. The results are shown for all
compounds (5,388) and for the subsets (1,294) meeting the stability criteria.
Inverse structures tend to occur when $\chi^X - \chi^Y < 0$, in accordance with
Burch's rule. This pattern remains even after considering the stability
criteria.

Additionally, when stability criteria are applied, stable inverse structures
generally have $X$ and $Y$ elements with similar covalent radii. This
observation aligns with the expectation that $X$ and $Y$ elements should be
similar in size in inverse structures, as their positions are interchanged
compared to regular structures. The region defined by $\chi^X - \chi^Y <
	0.15$~eV and $|r^X_{\mathrm{cov}} - r^Y_{\mathrm{cov}}| < 0.20$ \si{\angstrom}
is highlighted in green, marking a range where structural preferences are
distinct. Among the 490 stable inverse compounds identified, 390 (80\%) fall
within this region. However, the number of regular compounds satisfying these
criteria is comparable to inverse compounds, indicating that these criteria are
robust necessary conditions but not sufficient. Conversely, for regular
structure prediction, the criteria exhibit high precision (92\%) but lower
recall (75\%).

\subsection{Tetragonal distortion}

\begin{figure}[tb]
	\flushleft
	\includegraphics[width=0.48\textwidth]{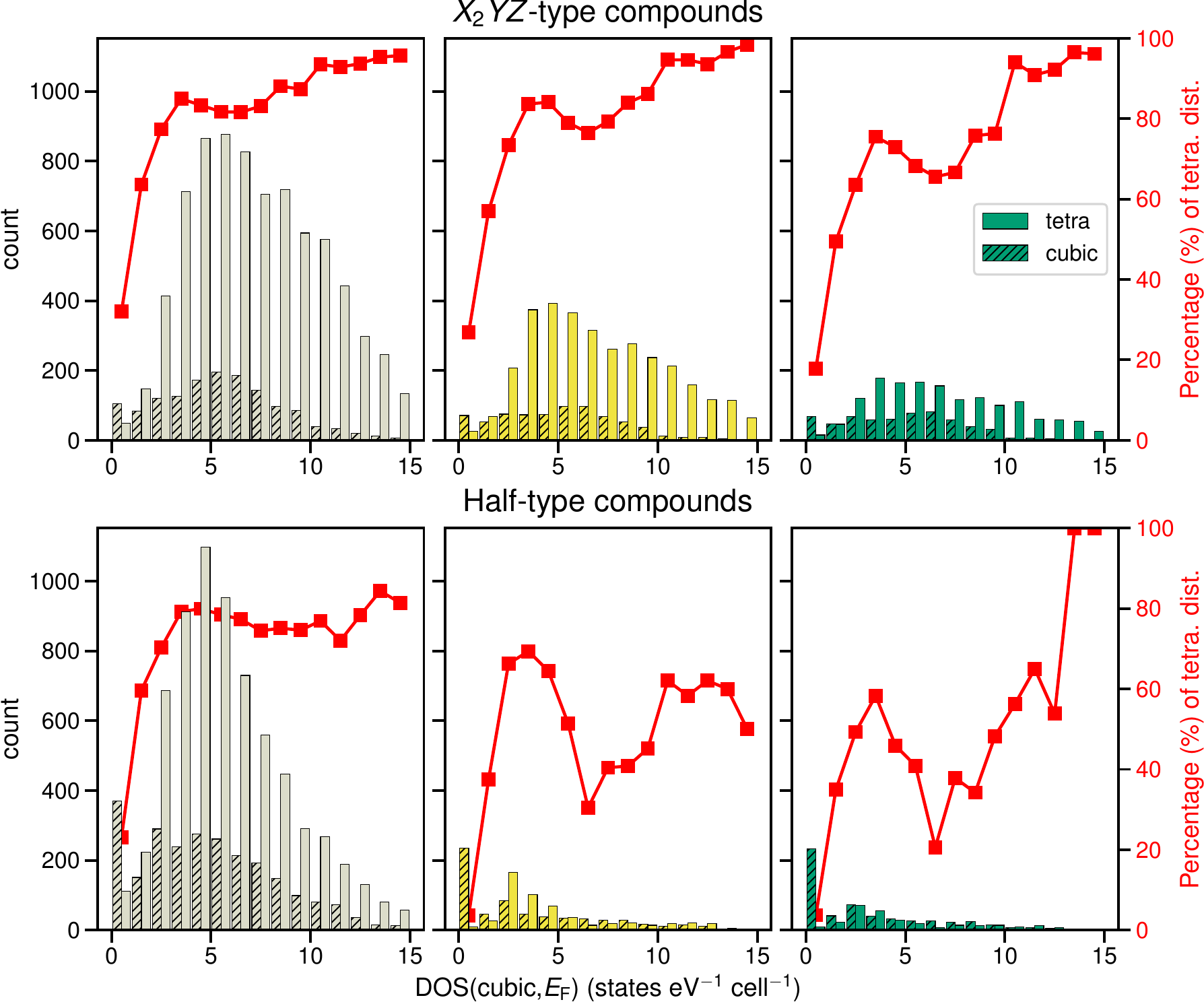}
	\caption{\label{fig:distor}
		Distribution of cubic and tetragonal Heusler compounds sorted into bins
		based on the density of states (DOS) values at the Fermi level in the
		cubic phase. The red curve represents the percentage of tetragonal
		distortion for each bin. The distributions are shown for all compounds
		(left panels), compounds meeting thermodynamic stability criteria
		(middle panels), and compounds additionally satisfying the dynamical
		stability criterion (right panels).}
\end{figure}

Heusler compounds typically crystallize in either the cubic or tetragonal phase.
Our \gls{htp} computational study revealed that 7,959 $X_2YZ$-type and 6,909
half-Heusler compounds exhibit tetragonal distortion in their ground state. The
primary factor contributing to tetragonal distortion is commonly thought to be
the peaks in the density of states (DOS) near the Fermi level (E$_F$) in the
cubic phase DOS(cubic, E$_F$) \cite{faleev_origin_2017}. This conclusion was
supported by the previous high-throughput studies on several hundred $X_2YZ$
compounds containing Fe, Co, and
Ni~\cite{winterlik_design_2012,wollmann_magnetism_2015,faleev_origin_2017,
	matsushita_large_2017}. The same conclusion is drawn from our \gls{htp} result
spanning significantly broader elemental space. The distribution of $X_2YZ$ type
compounds up to DOS(cubic, E$_F$)=15 eV$^{-1}$ is shown in Fig.\ref{fig:distor}.
The counts of cubic and tetragonal Heusler compounds are sorted into bins based
on their DOS(cubic,E$_F$) values. The red curve represents the probability of
tetragonal distortion in each bin. The probability of distortion increases as
DOS(cubic, E$_F$) rises. When DOS(cubic, E$_F$) exceeds 3 eV$^{-1}$, the
probability of tetragonal distortion is greater than 80\%. The same analysis was
extended to Half-Heusler compounds, and a similar behavior was found. When
DOS(cubic, E$_F$) exceeds 3 eV$^{-1}$, the probability of tetragonal distortion
is greater than 70\%.

It is worth noting that the inclusion of stability criteria alters the
distributions, as shown in the middle and right panels of Fig.~\ref{fig:distor}.
For half-Heusler compounds, tetragonal distortion is observed in 74\% of all
compounds, but this drops to 31\% when only phonon-stable compounds are
considered. The low ratio in stable compounds indicates that half-Heusler
compounds are less suitable for applications requiring tetragonal distortion,
such as materials with high \gls{mca}. Similarly, the occurrence of tetragonal
distortion in $X_2YZ$ compounds decreases from 85\% in all compounds to 72\%
among phonon-stable compounds.

\section{Conclusion}
In this study, we conducted a comprehensive high-throughput stability analysis
of 9,072/9,072/324/9,396 compositions of regular/inverse/$X_3Z$/half-Heusler
compounds in both cubic and tetragonal phases, considering various magnetic
configurations. In total, 106,235 structures in ground states and metastable
states were identified. By applying stability criteria to ground states based on
formation energy, Hull distance, and phonon stability, we identified 1,898
regular, 1,192 inverse, 81 $X_3Z$, and 840 half-Heusler stable compounds. Among
these, 1,356 compounds are magnetic, and 631 compounds exhibit \gls{tc} above
300~K, making them promising candidates for further experimental and theoretical
exploration of new functional materials. Notably, we identified 47 low-moment
\gls{fim} systems. The spin polarization and anomalous Hall/Nernst conductivity
were calculated to evaluate their potential applications in spintronics and
energy harvesting.

We validated the \gls{tc} calculation method within the mean-field approximation
against experimental data and found that a simple correction factor of
0.85 to the \Acrfull{fp}-based \gls{tc} provides a good agreement with
experimental results. Also, we validated \textit{ab initio} calculation
stability criteria alongside experimental data from the \gls{icsd}, suggesting
optimized criteria for achieving higher recall or accuracy, depending on
application needs. Our results demonstrated that the inclusion of phonon
stability significantly narrows the pool of viable candidates, emphasizing its
critical role. We expect these \gls{tc} calibration methods and refined
stability criteria to be valid to other types of Heusler compounds, such as
all-d and quaternary types, as well as similar magnetic systems.

Analysis of the \gls{htp} data revealed linear relationships between \gls{tc}
and $\sum_{i} |\bm{m}_{i}|$ in 14 Heusler compound systems, which can be used to
do rough estimation of \gls{tc}. We also identified correlations
between stability and atomic properties, such as atomic radius and ionization
energy. For regular/inverse preference in $X_2YZ$ compounds, our findings
aligned with the empirical Burch's rule and further indicated that inverse
structures are more likely when the $X$ and $Y$ elements have similar covalent
radii, although this condition is not sufficient to guarantee an inverse
structure formation. In terms of tetragonal distortion, we confirmed a strong
correlation between tetragonal distortion and a high density of states at the
Fermi level in the cubic phase in $X_2YZ$ compounds, in agreement with the previous
work, and identified a similar correlation in half-Heusler compounds.

Overall, these insights extend our understanding of stability and structural
preferences in Heusler compounds, providing a foundation for more efficient
material discovery in this and related categories. By refining the criteria for
stability and identifying key atomic correlations, this work contributes to the
development of next-generation functional materials. The comprehensive dataset
produced in this work, comprising 106,235 entries of the crystal structures,
magnetic moments, and the corresponding energies, 8,180 entries of phonons,
1,356 entries of \gls{tc}, and 355 entries of \gls{ahc}/\gls{anc}, is
available at \cite{dxmag_heuslerdb}. This database will be useful not only for
querying the stability of Heuslers but also for developing machine-learning
models. Development and application of such a machine learning model based on
our database, aiming to explore more complicated functional Heuslers
efficiently, are ongoing and will be present elsewhere.

\section*{Declaration of competing interest}
The authors declare that they have no known competing financial interests or
personal relationships that could have appeared to influence the work reported
in this paper.

\section*{Acknowledgments}
This study was supported by MEXT Program: Data Creation and Utilization-Type
Material Research and Development Project (Digital Transformation Initiative
Center for Magnetic Materials) Grant Number JPMXP112271550 and as ``Program for
Promoting Researches on the Supercomputer Fugaku'' (Data-Driven Research Methods
Development and Materials Innovation Led by Computational Materials Science,
JPMXP1020230327). This study used computational resources of supercomputer
Fugaku provided by the RIKEN Center for Computational Science (Project ID:
hp240223), the computer resources provided by ISSP, U-Tokyo under the program of
SCCMS, and the computer resources at NIMS Numerical Materials Simulator.

\bibliography{ref.bib}
\bibliographystyle{apsrev4-2}

\end{document}